\documentclass[a4paper]{llncs}
\makeatletter
\renewcommand*{\thetable}{\arabic{table}}
\renewcommand*{\thefigure}{\arabic{figure}}
\let\c@table\c@figure
\makeatother

\usepackage[inline]{enumitem}
\usepackage[noautolanguage,sepfour]{numprint}
\usepackage[pdftex]{graphicx}
\usepackage{subcaption}

\newif\iflong
\longtrue
\newif\ifminted
\mintedfalse

\usepackage{times}
\usepackage{fontawesome}
\newcommand{\checkOK}[1][green]{\text{\color{#1}\faCheck}}
\newcommand{\checkFail}[1][red]{\text{\color{#1}\faClose}}

\usepackage[T1]{fontenc}
\usepackage{hyperref}

\usepackage{amsmath,amssymb}
\usepackage{array,multirow}
\usepackage{xcolor}
\usepackage{xspace}
\usepackage{ifthen}
\newcommand{\custompar}[1]{\parskip 2pt \textbf{\textit{#1}}}

\newcommand{\mub}{\texttt{$\mu$gie}\xspace}
\newcommand{\mugie}{\mub}
\newcommand{\smt}{\textsc{smt}\--\textsc{lib}\xspace}
\newcommand{\btw}{\texttt{b2w}\xspace}

\usepackage[T1]{fontenc}
\usepackage[scaled=0.81]{beramono} 

\definecolor{boogieKW}{RGB}{27,158,119}
\definecolor{boogieCC}{RGB}{128,128,128}

\usepackage{tikz}
\usetikzlibrary{shapes,arrows,positioning,calc,fit,intersections,through}
\definecolor{toolcol}{HTML}{af8dc3}
\definecolor{arrowcol}{HTML}{808080}
\definecolor{bgcol}{HTML}{f7f7f7}

\usepackage{pgfkeys}
\newcommand{\setVal}[2]{\pgfkeyssetvalue{#1}{#2}}
\newcommand{\getVal}[1]{\pgfkeysvalueof{#1}}

\setVal{muPP}{P$_1$}
\setVal{muS1}{\ref{m:S1}}
\setVal{muS5}{\ref{m:S5}}
\setVal{muS6}{\ref{m:S6}}
\setVal{muS7}{\ref{m:S7}}
\setVal{muL1}{\ref{m:L1}}
\setVal{muL2}{\ref{m:L2}}
\setVal{muL4}{\ref{m:L4}}
\setVal{muL5}{\ref{m:L5}}
\setVal{muL6}{\ref{m:L6}}
\setVal{muL8}{\ref{m:L8}}
\setVal{muG1}{\ref{m:G1}}
\setVal{muG2}{\ref{m:G2}}

\ifminted
\usepackage{minted}
\newminted[boogie]{boogie}{mathescape,escapeinside=\#\#}
\newmintinline[B]{boogie}{}

\else
\usepackage{listings}
\usepackage{boogie}
\fi

\usepackage{booktabs}
\usepackage{changepage}
\usepackage{xparse}
\usepackage[]{algorithm2e}
\NewDocumentCommand{\lr}{m o}{ln.~\ref{ln:#1}\IfNoValueTF{#2}{}{--\ref{ln:#2}}}

\iflong
\makeatletter
\newcommand\pparagraph{\@startsection{paragraph}{4}{\z@}%
  {-1\p@ \@plus -0.5\p@ \@minus -0.5\p@}%
  {-0.5em \@plus -.22em \@minus -0.1em}%
  {\normalfont\normalsize\bfseries}}
\makeatother
\else
\newcommand\pparagraph[1]{\textbf{#1}}
\fi

\begin{document}

\title{Robustness Testing of Intermediate Verifiers}

\author{YuTing Chen \and Carlo A.\ Furia}

\institute{
Chalmers University of Technology, Sweden\\
\email{yutingc@chalmers.se} $\quad$ \url{bugcounting.net}
}

\maketitle

\begin{abstract}
Program verifiers are not exempt from the bugs that affect nearly every piece of software.
In addition, they often exhibit \emph{brittle} behavior:
their performance
changes considerably with details of how the input program is expressed---details 
that should be irrelevant, such as the order of independent declarations.
Such a lack of robustness frustrates users who have to spend considerable time figuring out 
a tool's idiosyncrasies before they can use it effectively.

This paper introduces a technique to detect lack of robustness of program verifiers;
the technique is lightweight and fully automated, as it is based on \emph{testing} methods (such as mutation testing and metamorphic testing).
The key idea is to generate many simple variants of a program that initially passes verification.
All variants are, by construction, equivalent to the original program;
thus, any variant that fails verification indicates lack of robustness in the verifier.

We implemented our technique in a tool called \mub, which operates on programs written in the popular Boogie language for verification---used as intermediate representation in numerous program verifiers.
Experiments targeting 135 Boogie programs indicate that brittle behavior occurs fairly frequently (16 programs) and is not hard to trigger.
Based on these results, the paper discusses the main sources of brittle behavior and suggests means of improving robustness.
\end{abstract}

\section{Introduction}
\label{sec:intro}

Automated program verifiers have become complex pieces of software;
inevitably, they contain bugs that make them misbehave in certain conditions.
\emph{Verification tools need verification too.}

In order to apply verification techniques to program verifiers, we have to settle on the kind of (correctness) properties to be verified.
If we simply want to look for basic \emph{programming errors}---such as memory allocation errors, or parsing failures---the usual verification\footnote{In this paper, the term ``verification'' also designates \emph{validation} techniques such as testing.} techniques designed for generic software---from random testing to static analysis---will work as well on program verifiers.
Alternatively, we may treat a program verifier as a \emph{translator} that encodes the semantics of a program and specification language into purely logic constraints---which can be fed to a generic theorem prover.
In this case, we may pursue a cor\-rect-by-con\-struc\-tion approach that checks that the translation preserves the intended semantics---as it has been done in few milestone research achievements~\iflong\cite{compcert,VeryAnalyzer}\else\cite{compcert}\fi.

There is a third kind of analysis, however, which is peculiar to automated program verifiers that aim at being sound.
Such tools input a program complete with specification and other auxiliary annotations,
and output either ``\checkOK~\textsc{success}'' or ``\checkFail~\textsc{failure}''.
Success means that the verifier proved that the input program is correct; 
but failure may mean that the program is incorrect or, more commonly, that the verifier needs more information to verify the program---such as more detailed annotations.
This asymmetry between ``verified'' and ``don't know'' is a form of incompleteness, which is inevitable for sound verifiers that target expressive, undecidable program logics.
Indeed, using such tools often requires users to become acquainted with the tools' idiosyncrasies, developing an intuition for what kind of information, and in what form, is required for verification to succeed.
To put it in another way, program verifiers may exhibit \emph{brittle, or unstable, behavior}: 
tiny changes of the input program that ought to be inconsequential 
have a major impact on the effectiveness achieved by the program verifier.
For instance, \autoref{sec:example} details the example of a small program that passes or fails verification just according to the relative order of two unrelated declarations. % in the input program.
Brittle behavior of this kind compromises the usability of verification tools. %---especially for beginner users who are easily confused by the inconsistent feedback they receive. %d from the verifier.

In this work, we target this kind of \emph{robustness (stability) analysis} of program verifiers.
We call an automated verifier \emph{robust} if its behavior is not significantly affected by small changes in the input that should be immaterial. 
A verifier that is not robust is \emph{brittle} (unstable): it depends on idiosyncratic features of the input.
Using brittle verifiers can be extremely frustrating: the feedback we get as we try to develop a verified program incrementally is inconsistent, and we end up running in circles---trying to fix nonexistent errors or adding unnecessary annotations.
Besides being a novel research direction for the verification of verifiers,
identifying brittle behavior has the potential of helping develop more robust tools that are ultimately more usable.

More precisely, we apply \emph{lightweight verification techniques} based on \emph{testing}. 
Testing is a widely used technique that cannot establish correctness but is quite effective at findings bugs.

The goal of our work is to automatically generate tests that reveal brittleness. 
Using the approach described in detail in \autoref{sec:how}, 
we start from a \emph{seed}: a program that is correct and can be verified by an automated verifier.
We \emph{mutate} the seed by applying random sequences of predefined mutation operators.
Each mutation operator captures a simple variation of the way a program is written that \emph{does not change its semantics}; for example, it changes the order of independent declarations. %, or renames a variable.
Thus, every mutant is a \emph{metamorphic transformation}~\cite{metamorphic} of the seed---and equivalent to it.
If the verifier \emph{fails} to verify a mutant we found a bug that exposes brittle behavior: 
seed and mutant differ only by small syntactic details that should be immaterial, 
but such tiny details impact the verifier's effectiveness in checking a correct program.

While our approach to \emph{robustness testing} is applicable in principle to any automated program verifier, the mutation operators depend to some extent on the semantics of the verifier's input language, as they have to be semantic preserving.
To demonstrate robustness testing in practice, we focus on the Boogie language~\cite{TIB2}.
Boogie is a so-called \emph{intermediate verification language}, 
combining an expressive program logic and a simple procedural programming language,
which is commonly used as an intermediate layer in many verification tools.
Boogie's popularity makes our technique (and our implementation) immediately useful to a variety of researchers and practitioners.

As we describe in \autoref{sec:how}, we implemented robustness testing for Boogie 
in a tool called \mub. 
In experiments described in \autoref{sec:experiments}, we ran \mub on 135 seed Boogie programs, generating and verifying over \numprint{87000} mutants.
The mutants triggered brittle behavior in 16 of the seed programs;
large, feature-rich programs turned out to be particularly brittle, to the point where several different mutations were capable of making Boogie misbehave.
As we reflect in \autoref{sec:discussion}, our technique for robustness testing can be a useful complement to traditional testing techniques, and it can help buttress the construction of more robust, and thus ultimately more effective and usable, program verifiers.

\custompar{Tool availability.}
The tool \mugie, as well as all the artifacts related to its experimental evaluation, 
are publicly available~\cite{toolSite}.
\iflong\else
A few additional details about the experiments are available in a longer version of this paper~\cite{RobustnessIVL-TR-20180509}.
\fi

\section{Motivating Example}
\label{sec:example}

Let's see a concrete example of how verifiers can behave brittlely.
\autoref{lst:live-verified} shows a simple Boogie program consisting of five declarations, each listed on a separate numbered line.

\begin{figure}[!htb]
\centering
\begin{boogie}[numbers=left,xleftmargin=20mm]
function h(int) returns (int);#\label{ln:hdecl}#
axiom (forall x, y: int :: x > y ==> h(x) > y);#\label{ln:hdef}#
const a: [int] int;#\label{ln:adecl}#
axiom (forall i: int :: 0 <= i ==> a[i] < a [i +$\,$1]);#\label{ln:adef}#
procedure p(i: int) returns (o: int) #\newline# $\quad$requires i >=$\,$0; ensures o > $\,$a[i]; { o := $\,$h(a[i +$\,$1]); }#\label{ln:pdef}#
\end{boogie}
\caption{A correct Boogie program that exposes the brittleness of verifiers: changing the order of declarations may make the program fail verification.}
\label{lst:live-verified}
\end{figure}

The program introduces an integer function \B{h} (\lr{hdecl}), whose semantics is partially axiomatized (\lr{hdef}); a constant integer map \B{a} (\lr{adecl}), whose elements at nonnegative indexes are sorted (\lr{adef}); and a procedure \B{p} (\lr{pdef}, spanning two physical lines in the figure)---complete with signature, specification, and implementation---which returns the result of applying \B{h} to an element of \B{a}.
Never mind about the specific nature of the program; we can see that procedure \B{p} is correct with respect to its specification: \B{a[i + 1] > a[i]} from the axiom about \B{a} and \B{p}'s precondition, and thus \B{h(a[i + 1]) > a[i] == o} from the axiom about \B{h}.
Indeed, Boogie successfully checks that \B{p} is correct.

There is nothing special about the order of declarations in \autoref{lst:live-verified}---after all, ``the order of the declarations in a [Boogie] program is immaterial''~\cite[Sec.~1]{TIB2}.
A different programmer may, for example, put \B{a}'s declarations before \B{h}'s.
In this case, surprisingly, Boogie fails verification warning the user that \B{p}'s postcondition may not hold.\iflong\footnote{The first author found out this at the most inappropriate of times---during a live demo!}\fi

A few more experiments show that there's a fair chance of running into this kind of brittle behavior.
Out of the $5! = 120$ possible permutations of the 5 declarations in \autoref{lst:live-verified}---each
an equivalent version of the program---Boogie verifies exactly half, and fails verification of the other half.
We could not find any simple pattern in the order of declarations (such as ``line $x$ before line $y$'') that predicts whether a permutation corresponds to a program Boogie can verify.

\iflong
The \smt~\cite{SMTlib2} files generated by Boogie (encoding the program's correctness conditions in a format understood by SMT solvers) also only differ by the order of declarations and assertions---with additional complications due to the fact that declarations have to come before usage in \smt, and the VCs also include auxiliary functions generated by Boogie.
The interplay of Boogie and Z3 determines the brittle behavior we observe in this example: even if it uses a generic format, Boogie's \smt encoding uses plenty of Z3-specific options, but is not always robust in the way it interacts with the solver.
\fi
To better understand whether other tools' SMT encodings may be less brittle than Boogie's,
we used \btw~\cite{b2w} to translate all 120 permutations of \autoref{lst:live-verified} to WhyML---the input language of the Why3 intermediate verifier~\cite{FilliatreP13}.
Why3 successfully verified all of them---using Z3 as SMT solver, like Boogie does---which suggests that some features of Boogie's encoding (as opposed to Z3's capabilities) are responsible for the brittle behavior on the example.

Such kinds of brittleness---a program switching from verified to unverified based on changes that should be inconsequential---can greatly frustrate users, and in particular novices who are learning the ropes and may get stuck looking for an error in a program that is actually correct---and could be proved so if definitions were arranged in a slightly different way.
Since brittleness hinders scalability to projects of realistic size, it can also be a significant problem for advanced users; for example, the developers behind the Ironclad Apps~\cite{Ironclad} and IronFleet~\cite{IronFleet} projects reported\footnote{By an anonymous reviewer of FM 2018.} that ``solvers' instability was a major issue'' in their verification efforts.

\begin{figure}[!hb]
\begin{adjustwidth}{-30mm}{-30mm}
\centering
\begin{tikzpicture}[
  toolbox/.style={rectangle,minimum width=20mm,minimum height=5mm, very thick,rounded corners=2mm,font=\footnotesize,draw=boogieKW,fill=boogieKW,text=white},
  databox/.style={font=\footnotesize,minimum width=16mm},
  outbox/.style={font=\footnotesize,minimum width=6mm},
  align=center
  ]
  \matrix[row sep=1mm,column sep=7mm] {
  & & \node (m1) [databox] {Mutant $m_1$}; 
    & \node (v1) [toolbox] {Verifier $t$};
    & \node (o1) [outbox] {\checkOK};
  \\
  & & \node (m2) [databox] {Mutant $m_2$}; 
    & \node (v2) [toolbox] {Verifier $t$};
    & \node (o2) [outbox] {\checkOK};
  \\
  \node (seed) [databox] {Seed: program $s$: \\$t(s) = \checkOK$};
    &  \node (morph) [toolbox,draw=toolcol,minimum height=10mm,fill=toolcol,font=\footnotesize\bfseries] {Mutation\\generator \mub};
    & \node (m22) [databox] {$\vdots$}; 
    & \node (v22) [toolbox,draw=none,fill=none] {$\vdots$}; 
    & \node (o22) [outbox] {$\vdots$};
  \\
  & & \node (mk) [databox] {Mutant $m_{k}$}; 
    & \node (vk) [toolbox] {Verifier $t$};
    & \node (ok) [outbox] {\checkFail};
    & \node (found) [databox,fill=red!30!white,very thick] {Brittle behavior of $t$: \\ $t(m_k) \neq t(s)$};
  \\
  & & \node (mk2) [databox] {$\cdots$}; 
    & \node (vk2) [toolbox,draw=none,fill=none] {$\cdots$}; 
    & \node (ok2) [outbox] {\checkOK};
  \\
  & & \node (mn) [databox] {Mutant $m_N$};  
    & \node (vn) [toolbox] {Verifier $t$};
    & \node (on) [outbox] {\checkOK};
  \\
  };
  \begin{scope}[color=arrowcol,line width=1pt,round cap-latex',shorten >=3pt,shorten <= 4pt,every node/.style={font=\footnotesize}]
  \draw (seed) -- (morph);
   \foreach  \k in {1, 2, 22, k, k2, n} {
     \draw (morph) -- (m\k.west);
     \draw (m\k.east) -- (v\k);
     \draw (v\k) -- (o\k.west);
   }
   \draw (ok) -- (found);
  \end{scope}
\end{tikzpicture}
\end{adjustwidth}
\caption{\footnotesize How robustness testing of Boogie programs works.
We start with a correct program $s$ that some Boogie tool $t$ can successfully verify;
mutation generator \mub mutates $s$ in several different ways, generating many different \emph{mutants} $m_k$ equivalent to $s$;
each mutant undergoes verification with tool $t$; 
a mutant $m_k$ that \emph{fails verification} with $t$ exposes \emph{brittle behavior} of $t$ 
on the two equivalent correct programs $s \equiv m_k$.
}
\label{fig:workflow}
\end{figure}
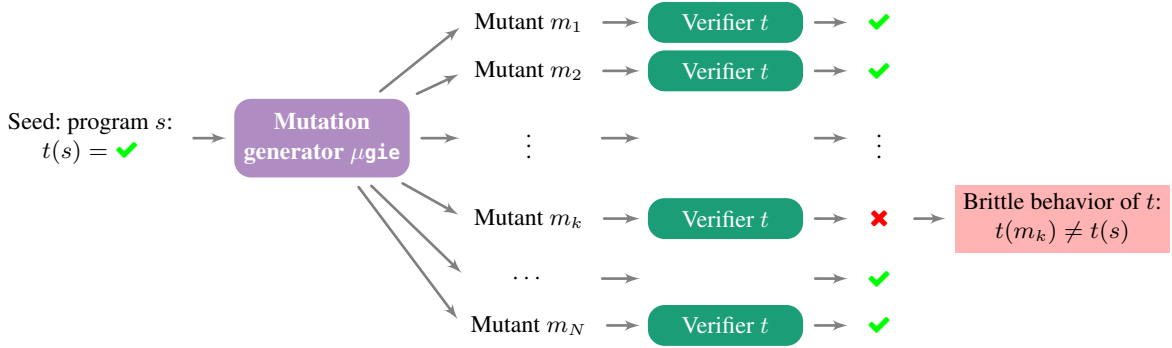

\section{How Robustness Testing Works}
\label{sec:how}

Robustness testing is a technique that ``perturbs'' 
a correct and verified program by introducing small changes, 
and observes whether the changes affect the program's verifiability.
The changes should be inconsequential, because they are designed 
not to alter the program's behavior or specification;
if they do change the verifier's outcome, we found lack of robustness.

While robustness testing is applicable to any automated program verifier,
we focus the presentation on the popular Boogie intermediate verification language.
Henceforth, a ``program'' is a program (complete with specification and other annotations) written in the Boogie language.
\autoref{fig:workflow} illustrates how robustness testing works at a high level;
the rest of the section provides details.

In general terms, testing requires to build a valid \emph{input}, feed it to the system under test, and compare the system's output with the \emph{expected} output---given by a testing \emph{oracle}.
Testing the behavior of a verifier according to this paradigm 
brings challenges that go beyond those involved in generating tests for general programs.
First, a verifier's input is a whole \emph{program}, complete with specification and other annotations (such as lemmas and auxiliary functions) for verification.
Second, robustness testing aims at exposing subtle inconsistencies in a verifier's output, and not basic programming errors---such as memory access errors, parsing errors, or input/output errors---that every piece of software might be subject to.
Therefore, we need to devise suitable strategies for \emph{input generation} and \emph{oracle generation}.

\subsection{Mutation Operators}

\pparagraph{Input generation.}
In order to expose brittleness of verifiers, we need to build complex input programs of significant size, complete with rich specifications and all the annotations that are necessary to perform automated verification.
While we may use grammar-based generation techniques~\cite{csmith} to automatically build syntactically correct Boogie programs, the generated programs would either have trivial specifications or not be semantically correct---that is, they would not pass verification.
Instead, robustness testing starts from a collection of \emph{verified programs}---the \emph{seeds}---and automatically generates simple, semantically equivalent variants of those programs.\footnote{\iflong\autoref{sec:results}\else\cite{RobustnessIVL-TR-20180509}\fi{} describes some experiments with seeds that \emph{fail} verification. Unsurprisingly, random mutations are unlikely to turn an unverified program into a verified one---therefore, the main paper focuses on using verified programs as seeds.}
This way, we can seed robustness testing with a variety of sophisticated verification benchmarks, and assess robustness on realistic programs of considerable complexity.

\pparagraph{Mutation operators.}
Given a seed $s$, robustness testing generates many variants $M(s)$ of $s$ by ``perturbing'' $s$.
Building on the basic concepts and terminology of mutation testing~\iflong\cite{mutation,mutationSurvey}\else\cite{mutationSurvey}\fi,\footnote{See \autoref{sec:related} for a discussion of how robustness testing differs from traditional mutation testing.}
we call \emph{mutant} each variant $m$ of a seed $s$ obtained by applying a random sequence of \emph{mutation operators}.

A mutation operator captures a simple syntactic transformation of a Boogie program;
crucially, mutation operators should \emph{not change a program's semantics} 
but only introduce equivalent or redundant information.
Under this fundamental condition, every mutant $m$ of a seed $s$ is equivalent 
to $s$ in the sense that $s$ and $m$ should both pass (or both fail) verification.
This is an instance of \emph{metamorphic testing}, where we transform between equivalent inputs so that the seed serves as an oracle to check the expected verifier output on all of the seed's mutants.

\begin{table}[!bt]
\centering
\scriptsize
\begin{tabular}{*{3}{p{0.32\textwidth}}}
\multicolumn{1}{c}{\textsc{structural}}
&
\multicolumn{1}{c}{\textsc{local}}
&
\multicolumn{1}{c}{\textsc{generative}}
\\
\midrule
\begin{enumerate}[label=S$_{\arabic*}$]
\item \label{m:S1} Swap any two declarations
\item \label{m:S5} Split a procedure definition into declaration and implementation
\item \label{m:S6} Move any declaration into a separate file (and call Boogie on both files)
\end{enumerate}
&
\begin{enumerate}[label=L$_{\arabic*}$]
\item \label{m:L1} Swap any two local variable declarations
\item \label{m:L2} Split a declaration of multiple variables into multiple declarations
\item \label{m:L4} Join any two preconditions into a conjunctive one
\item \label{m:L5} Join any two postconditions into a conjunctive one
\item \label{m:L6} Swap any two pre-/postcondition, intermediate assertion, or loop invariant clauses
\item \label{m:L8} Complement an {\B{if}} condition and switch its {\B{then}} and {\B{else}} branches
\end{enumerate}
&
\begin{enumerate}[label=G$_{\arabic*}$]
\item \label{m:G1} Add {\B{true}} as pre-/postcondition, intermediate assertion, or loop invariant clause
\item \label{m:G2} Remove a trigger annotation
\end{enumerate}
\end{tabular}
\caption{
Mutation operators of Boogie code in categories structural, local, and generative.
Operators do not change the semantics of the code they are applied to (except possibly \ref{m:G2}, which is used separately).}
\label{tab:transformations}
\end{table}

Based on our experience using Boogie and working around its brittle behavior, we designed the
mutation operators in \autoref{tab:transformations}, which exercise different language features:
\begin{description}
\item[Structural] mutation operators change the overall structure of top-level declarations---by changing their relative order (\ref{m:S1}), separating declarations and implementations (\ref{m:S5}), and splitting into multiple files (\ref{m:S6}).
\item[Local] mutation operators work at the level of procedure bodies---by changing the relative order of or splitting on multiple lines local variable declarations (\ref{m:L1} and \ref{m:L2}), 
merging two pre- or postcondition clauses \B{x} and \B{y} into a conjunctive clause \B{x && y} (\ref{m:L4} and \ref{m:L5}), 
changing the relative order of assertions of the same program element (\ref{m:L6}),
and permuting the \B{then} and \B{else} branches of a conditional (\ref{m:L8}).
\item[Generative]  mutation operators alter redundant information---by adding trivial assertions (\ref{m:G1}), and removing quantifier instantiation suggestions (``triggers'' in \ref{m:G2}).
\end{description}

We stress that our mutation operators
do not alter the semantics of a Boogie program according to the language's specification~\cite{TIB2}:
in Boogie, 
the order of declarations is immaterial (\ref{m:S1}, \ref{m:L1}, \ref{m:L2});
a procedure's implementation may be with its declaration or be separate from it (\ref{m:S5});
multiple input files are processed as if they were one (\ref{m:S6});
multiple specification elements are implicitly conjoined, and their relative order does not matter (\ref{m:L4}, \ref{m:L5}, \ref{m:L6});
a conditional's branches are mutually exclusive (\ref{m:L8});
and 
\B{true} assertions are irrelevant since Boogie only checks partial correctness (\ref{m:G1}).

\pparagraph{Triggers.}
\ref{m:G2} is the only mutation operator that may alter the semantics of a Boogie program in practice: while triggers are suggestions on how to instantiate quantifiers, they are crucial to guide SMT solvers and increase stability in practice~\cite{triggerSelection,blt}.
Therefore, \textbf{we do not consider \ref{m:G2} semantics-preserving}; our experiments only apply \ref{m:G2} in a separate experimental run to give an idea of its impact in isolation.

More mutation operators are possible, but the selection in \autoref{tab:transformations} should strike a good balance between effectiveness in setting off brittle behavior and feasibility of studying the effect of each individual operator in isolation.

\iflong
\pparagraph{Example.}
In the example of \autoref{sec:example},
the program in \autoref{lst:live-verified} is a possible seed---a correct Boogie program that verifies.
Applying mutation operator \ref{m:S1} twice---first 
to lines \autoref{ln:hdecl}, \autoref{ln:adecl}, and then
to lines \autoref{ln:hdef}, \autoref{ln:adef}---generates a mutant
where \B{a}'s declarations come before \B{h}.
As discussed in \autoref{sec:example}, this mutant fails verification even if it 
is equivalent to the seed.
\fi

\begin{algorithm}[!bt]
\DontPrintSemicolon
\SetCommentSty{textnormal}
\SetDataSty{textnormal}

\SetKwInOut{Input}{input}
\SetKwInOut{Output}{output}
\SetKwProg{break}{break}{}{}
\SetKwData{attempts}{attempts}
\SetKwData{MAX}{MAX\_ATTEMPTS}
  \Input{seed program $s$}
  \Input{weight $w(o)$ for each mutation operator $o$}
  \Input{number of mutants $N_M$}
  \Output{set of mutants $M$ of $s$}
  \BlankLine
  \BlankLine  
  $M \gets \{ s \}$ \tcp*{initialize pool of mutants to seed}
  attempts $\gets 0$ \tcp*{number of main loop iterations}

  \While(\tcp*[f]{repeat until $N_M$ mutants are generated}){$|M| < N_M$}{  
    \If{\attempts $>$ \MAX}{
      \break{}{}
    }{
      $p \gets$ any program in $M$\;
      $o \gets$ any mutation operator  \tcp*{draw with probability $w(o)$}
      $m \gets o(p)$  \tcp*{apply mutation operator $o$ to $p$}
      $M \gets M \cup \{m\}$ \tcp*{add $m$ to pool $M$}
    }
  attempts $\gets$ attempts $+$ 1
  }
  \Return{$M$}
  \caption{Mutant generation algorithm}
  \label{alg:mugen}
\end{algorithm}

\subsection{Mutation Generation}

Given a seed $s$, the generation of mutants repeatedly draws random mutation operators and applies them to $s$, or to a previously generated mutant of $s$, until the desired number $N_M$ of mutants is reached.

Alg.~\ref{alg:mugen} shows the algorithm to generate mutants.
The algorithm maintains a pool $M$ of mutants, which initially only includes the seed $s$.
Each iteration of the main generation loop proceeds as follows:
\begin{enumerate*}
\item pick a random program $p$ in the pool $M$;
\item select a random mutation operator $o$;
\item apply $o$ to $p$, giving mutant $m$;
\item add $m$ to pool $M$ (if it is not already there).
\end{enumerate*}

Users can bias the random selection of mutation operators by assigning a weight $w(o)$ to each mutation operator $o$ in \autoref{tab:transformations}:
the algorithm draws an operator with probability proportional to its weight,
and operators with zero weight are never drawn.

Besides the mutation operator selection, there are two other passages of the algorithm where random selection is involved:
a program $p$ is drawn uniformly at random from $M$;
and applying an operator $o$ selects uniformly at random program locations where $o$ can be applied.
For example, if $o$ is \ref{m:S1} (swap two top-level declarations), applying $o$ to $p$ involves randomly selecting two top level declarations in $p$ to be swapped.

Any mutation operator can generate only finitely many mutants;
since the generation is random, it is possible that a newly generated mutant is identical to one that is already in the pool.
In practice, this is not a problem as long as the seed $s$ is not too small or the enabled operators too restrictive (for example, \ref{m:S5} can only generate $2^D$ mutants, where $D$ is the number of procedure definitions in $s$).
The generation loop has an alternative stopping conditions that gives up after 
MAX\_ATTEMPTS
iterations that have failed to generate enough distinct mutants.

\pparagraph{Robustness testing.}
After generating a set $M(s)$ of mutants of a seed $s$,
robustness testing runs the Boogie tool
on each mutant in $M(s)$.
If Boogie can verify $s$ but fails to verify any mutant $m \in M(s)$,
we have found an instance of \emph{brittle behavior}:
$s$ and $m$ are equivalent by construction,
but the different form in which $m$ is expressed trips up Boogie
and makes verification fail on an otherwise correct program.

\subsection{Implementation}
\label{sec:implementation}

We implemented robustness testing as a commandline tool \mub (pronounced ``moogie''). 
\mub implements in Haskell the mutation generation Alg.~\ref{alg:mugen}, 
and extends parts of Boogaloo's front-end~\cite{boogaloo} for parsing and typechecking Boogie programs.

\iflong
Each mutation operator is implemented---using the \verb|lens| Haskell package~\cite{haskellLens}---as function from Boogie programs to Boogie programs;\footnote{Precisely, the input domain also includes a random seed to select the program elements to which the operator is applied.}
With this design, adding new mutation operators or changing their order of application is straightforward as it just uses Haskell's function composition.
Upon terminating, \mugie outputs each mutant as a separate file (or files, if \ref{m:S6} is applied), and annotates each file with a comment header indicating the seed and the sequence of mutation operators that were applied to generate the mutant from the seed.
\fi

\iflong
\pparagraph{Pretty printing.}
\mub represents programs by their abstract syntax trees,
which are then rendered using concrete syntax when all mutants have been generated.
Such a pretty printing may introduce small syntactic changes---when the same abstract syntax can be rendered using different syntactic forms.
Besides obvious changes in white spaces, line breaks, and comments (which are removed),
\mugie's pretty printer may introduce one normalizing change: 
functions and procedures without return clause in their signature get an explicit \B{returns()} clause.
To further ensure no implicit transformation is added during pretty printing, we also checked that pretty printing has no effect on the behavior of Boogie programs (that is, a pretty-printed seed verifies iff the original seed also verifies).
\fi

\section{Experimental Evaluation}
\label{sec:experiments}

Robustness testing was initially motivated by our anecdotal experience using intermediate verifiers.
To rigorously assess to what extent they are indeed brittle, and whether robustness testing can expose their brittleness, we conducted an experimental evaluation using \mugie. 
This section describes design and results of these experiments.

\subsection{Experimental Design}

A run of \mugie inputs a \emph{seed} program $s$ and outputs a number of metamorphic mutants of $s$, which are then verified with some tool $t$ (see \autoref{fig:workflow}).

\pparagraph{Seed selection.}
We prepared a curated collection of seeds by selecting Boogie programs from several different sources, with the goal of having a diverse representation of how Boogie may be used in practice.
Each example belongs to one of six groups according to its origin and characteristics; \autoref{tab:loc} displays basic statistics about them.
Group \textbf{A} contains basic \textbf{A}lgorithms (search in an array, binary search trees, etc.) implemented directly in Boogie in our previous work~\cite{FMV-CSUR14}; 
these are relatively simple, but non-trivial, verification benchmarks.
Group \textbf{T} is a different selection of mainly algorithmic problems (bubble sort, Dutch flag, etc.) included in Boogie's distribution \textbf{T}ests.
Group \textbf{E} consists of small \textbf{E}xamples from our previous work~\cite{blt} that target the impact of different trigger annotations in Boogie.
Group \textbf{S} collects large Boogie programs that we generated automatically from fixed, repetitive structures (for example, nested conditionals); 
in previous work~\cite{blt} we used these programs to evaluate \textbf{S}calability.
Groups \textbf{D} and \textbf{P} contain Boogie programs automatically generated by the \textbf{D}afny~\cite{dafny} and Auto\textbf{P}roof~\cite{FNPT-STTT16} verifiers (which use Boogie as intermediate representation).
The Dafny and Eiffel programs they translate come from the tools' galleries of verification benchmarks~\cite{dafnyGallery,autoproofGallery}.
As we see from the substantial size of the Boogie programs they generate, Dafny and AutoProof introduce a significant overhead as they include axiomatic definitions of heap memory and complex types.
In all, we collected 135 seeds of size ranging from just 6 to over \numprint{8500} lines of Boogie code for a total of nearly \numprint{260000} lines of programs and specifications.

\begin{table}[!hbt]
\scriptsize
\begin{subtable}[c]{0.5\textwidth}
\centering
\setlength{\tabcolsep}{2.2pt}
\begin{tabular}{crrrrrrr}
\toprule
&& \multicolumn{5}{c}{\textsc{loc}} \\
\textsc{group} & \textsc{\# seeds} & \textsc{min} & \textsc{median} & \textsc{mean} & \textsc{max} & \textsc{total} \\ 
\midrule
A & \numprint{10} & \numprint{17} & \numprint{34} & \numprint{44} & \numprint{152} & \numprint{439} \\
D & \numprint{26} & \numprint{2000} & \numprint{4076} & \numprint{4465} & \numprint{8533} & \numprint{116101} \\
E & \numprint{10} & \numprint{13} & \numprint{18} & \numprint{23} & \numprint{49} & \numprint{230} \\
P & \numprint{30} & \numprint{986} & \numprint{1665} & \numprint{1911} & \numprint{5737} & \numprint{57330} \\
S & \numprint{51} & \numprint{6} & \numprint{126} & \numprint{1047} & \numprint{7286} & \numprint{67006} \\
T & \numprint{8} & \numprint{11} & \numprint{41} & \numprint{1662} & \numprint{7378} & \numprint{18283} \\
\midrule
\textsl{all} & \numprint{135} & \numprint{6} & \numprint{642} & \numprint{1718} & \numprint{8533} & \numprint{259389} \\
\bottomrule
\end{tabular}

\caption{\scriptsize 
Selection of Boogie programs used as seeds: for each \textsc{group}, the number of programs in that group (\textsc{\# seeds}), and their \textsc{min}imum, \textsc{median}, \textsc{mean}, \textsc{max}imum, and \textsc{total} size in non-blank non-comment lines of code. Row \textsl{all} summarizes measures over all groups.}
\label{tab:loc}
\end{subtable}
\hspace{7mm}
\begin{subtable}[c]{0.4\textwidth}
\centering
\setlength{\tabcolsep}{2.2pt}
\begin{tabular}{lrrr}
\toprule
\textsc{tool} & \textsc{commit} & \textsc{date} & \textsc{z3} \\
\midrule
\textsc{boogie} 4.1.1  &  \texttt{b2d448}  & 2012-09-18 & 4.1.1
\\
\textsc{boogie} 4.3.2  &  \texttt{97fde1}  & 2015-03-10 & 4.3.2
\\
\textsc{boogie} 4.4.1  &  \texttt{75b5be}  & 2015-11-19 & 4.4.1
\\
\textsc{boogie} 4.5.0  &  \texttt{63b360}  & 2017-07-06 & 4.5.0
\\
\bottomrule
\end{tabular}
\caption{\scriptsize 
Selection of Boogie versions used in the experiments.
For every version of the Boogie \textsc{tool}, the corresponding \textsc{commit} hash in Boogie's Git repository, the \textsc{date} of the commit, and the matching \textsc{z3} version.}
\label{tab:boogie-versions}
\end{subtable}
\caption{Boogie programs (``seeds'') and Boogie tool versions used in the experiments.}
\label{tab:basic-stats}
\end{table}

\pparagraph{Tool selection.}
In principle, \mugie can be used to test the robustness of any verifier that can input Boogie programs:
besides Boogie, tools such as Boogaloo~\cite{boogaloo}, Symbooglix~\cite{symbooglix}, and \texttt{blt}~\cite{blt}.
However, different tools target different kinds of analyses, and thus typically require different kinds of seeds to be tested properly and meaningfully compared.
To our knowledge, no tools other than Boogie itself  
support the full Boogie language, or are as mature and as effective as Boogie for sound verification (as opposed to other analyses, such as the symbolic execution performed by Boogaloo and Symbooglix) on the kinds of examples we selected.
We intend to perform a different evaluation of these tools using \mugie in the future, but for consistency and clarity we focus on the Boogie tool in this paper.

In order to understand whether Boogie's robustness has changed over its development history, our experiments include different versions of Boogie.
The Boogie repository is not very consistent in assigning new version numbers, nor does it tag specific commits to this effect.
As a proxy for that, we searched through the logs of Boogie's repository for commit messages that indicate updates to accommodate new features of the Z3 SMT solver---Boogie's standard and main backend.
For each of four major versions of Z3 (4.1.1, 4.3.2, 4.4.1, and 4.5.0), we identified the most recent commit that refers explicitly to that version (see \autoref{tab:boogie-versions});
for example, commit \verb|63b360| says ``Calibrated test output to Z3 version 4.5.0''.
Then, we call ``Boogie~$v$'' the version of Boogie at the commit mentioning Z3 version $v$, running Z3 version $v$ as backend.

To better assess whether brittle behavior is attributable to Boogie's encoding or to Z3's behavior, we included two other tools in our experiments:
CVC4 refers to the SMT solver CVC4 v.~1.5 inputting Boogie's SMT2 encoding of verification condition (the same input that is normally fed to Z3);
Why3 refers to the intermediate verifier Why3 v.~0.86.3 using Z3~4.3.2 as backend, and inputting WhyML translations of Boogie programs automatically generated by \btw~\cite{b2w}.

\begin{table}[!htb]
\begin{adjustwidth}{-10mm}{5mm}
\scriptsize
\setlength{\tabcolsep}{2pt}
\begin{tabular}{rlp{0.645\textwidth}}
\toprule
  & \textsc{definition} & \textsc{description} \\
\midrule
  $S$  &  &  set of all seeds \\
  $M_{O}(s)$ &  &  set of all mutants of seed $s$ (generated with mutation operators $O$) \\
  \midrule
  $S_t^{\checkOK}$
  & $\{ s \in S \mid t(s) \}$
  & seeds that pass verification with tool $t$
  \\
  $M_O(s)^{\checkFail}_t$
  & $\{ m \in M_O(s) \mid \neg t(m) \}$
  & mutants of seed $s$ that fail verification with tool $t$
  \\
  $S_t^{\checkOK \leadsto \checkFail}$
  & $\{ s \in S_t^{\checkOK} \mid |M_O(s)^{\checkFail}_t| > 0 \}$
  & passing seeds with at least one mutant failing with tool $t$
  \\
  $M_O(s)^{\infty}_t$
  & $\{ m \in M_O(s)^{\checkFail}_t \mid t(m) \text{ times out} \}$
  & failing mutants of seed $s$ that time out with tool $t$
  \\
  \midrule
  \textsc{\# pass}
  & $|S_t^{\checkOK}|$
  & number of seeds that pass verification with tool $t$
  \\
  \textsc{\# $\exists$fail}
  & $|S_t^{\checkOK \leadsto \checkFail}|$
  & number of verified seeds with at least one failing mutant with tool $t$
  \\
  \textsc{\% $\exists$fail}
  & $100\cdot|S_t^{\checkOK \leadsto \checkFail}|/|S_t^{\checkOK}|$
  & percentage of verified seeds with at least one failing mutant with tool $t$
  \\
  $\overline{\textsc{\% fail}}$
  & $100\cdot \textit{mean}_{s \in S_t^{\checkOK}} |M_O(s)^{\checkFail}_t|/|M_O(s)|$
  & average percentage of failing mutants per verified seed with tool $t$
  \\
  $\overline{\textsc{\% timeout}}$
  & $100\cdot\textit{mean}_{s \in S_t^{\checkOK}}|M_O(s)^{\infty}_t|/|M_O(s)|$
  & average percentage of timed out mutants per verified seed with tool $t$
  \\
  $\overline{\textsc{\% }\exists\textsc{fail}}$
  & $100\cdot\textit{mean}_{s \in S_t^{\checkOK \leadsto \checkFail}}|M_O(s)^{\checkFail}_t|/|M_O(s)|$
  & average percentage of failing mutants per verified seed with some failing mutants
  \\
\bottomrule
\end{tabular}
\end{adjustwidth}
\caption{Definitions and descriptions of the experimental measures reported in \autoref{tab:fail}.}
\label{tab:notation}
\end{table}

\pparagraph{Experimental setup.}
Each experiment has two phases: first, \emph{generate mutants} for every seed; then, \emph{run Boogie} on the mutants and check which mutants still verify.

For every seed $s \in S$ (where $S$ includes all 135 programs summarized in \autoref{tab:loc}),
we generate different \emph{batches} $M_O(s)$ of mutants of $s$ by enabling specific mutation operators $O$ in \mugie.
Precisely, we generate 12 different batches for every seed:% \footnote{Mutation generation always ran to completion, generating the target number of mutants before reaching the maximum number of attempts.}
\begin{description}
\item[$M_{*}(s)$] consists of 100 different mutants of $s$, generated by picking uniformly at random among all mutation operators in \autoref{tab:transformations} \textbf{except \getVal{muG2}} (that is, each mutation operator gets the same positive weight, and \getVal{muG2} gets weight zero);
\item[$M_J(s)${\normalfont,}] for $J$ one of the 11 operators in \autoref{tab:transformations}, consists of 50 different mutants of $s$ generated by only applying mutation operator $J$ (that is, $J$ gets a positive weight, and all other operators get weight zero).
\end{description}
Batch $M_*$ demonstrates the effectiveness of robustness testing with general settings; then,
the smaller batches $M_J$ focus on the individual effectiveness of one mutation operator at a time.
Operator \ref{m:G2} is only used in isolation (and not at all in $M_*$) since it may change the semantics of programs indirectly by guiding quantifier instantiation.

Let $t$ be a tool (a Boogie version in \autoref{tab:boogie-versions}, or another verifier).
For every seed $s \in S$,
we run $t$ on $s$
and on all mutants $M_O(s)$ in each batch.
For a run of $t$ on program $p$ (seed or mutant), we write $t(p)$ if $t$ verifies $p$ successfully; and $\neg t(p)$ if $t$ fails to verify $p$ (because it times out, or returns with failure).
Based on this basic data, we measure robustness by counting the number of verified seeds whose mutants fail verification: see the measures defined in \autoref{tab:notation} and the results described in detail in \autoref{sec:results}.
\iflong
As an additional check that the mutation operators do not change the semantics of programs, 
we ascertained that all mutants pass parsing and other syntactic checks.
\fi

\iflong
To reduce the running time in practical usage scenarios,
 one may stop verification of a seed's mutants as soon as one of them fails verification,
or even alternate seed mutation and verification steps to avoid generating mutants after the first failure.
In this paper's experiments, however, we decided to run verification exhaustively on every mutant in order to collect detailed information about the effectiveness of robustness testing.
\fi

\pparagraph{Running times.}
The experiments ran on a Ubuntu 16.04 LTS GNU/Linux box with Intel 8-core i7-4790 CPU at 3.6~GHz and 16~GB of RAM.
Generating the mutants took about 15 minutes for the batch $M_*$ and 10 minutes for each batch $M_J$.
Each verification run was given a timeout of 20 seconds, after which it was forcefully terminated by the scheduler of GNU \texttt{parallel}~\cite{gnuParallel}.
\iflong
For replicability, seeds and summary output 
of the experimental runs are available online~\cite{toolSite}.
\fi

\begin{table}[!tb]
\centering
\scriptsize
\setlength{\tabcolsep}{1.4pt}
\begin{tabular}{llccccccccccccccccc}
\toprule
& &
\multicolumn{6}{c}{\textsc{batch} $M_*$}
&
\multicolumn{11}{c}{\textsc{\# $\exists$fail of $M_J$}}
\\
\cmidrule(lr){3-8}
\cmidrule(lr){9-19}
\textsc{group} & \textsc{tool} & \textsc{\# pass} & \textsc{\# $\exists$fail} & \textsc{\% $\exists$fail} & $\overline{\textsc{\% fail}}$ & $\overline{\textsc{\% timeout}}$ & $\overline{\textsc{\% }\exists\textsc{fail}}$ & \textsc{\getVal{muS1}} & \textsc{\getVal{muS5}} & \textsc{\getVal{muS6}} & \textsc{\getVal{muL1}} & \textsc{\getVal{muL2}} & \textsc{\getVal{muL4}} & \textsc{\getVal{muL5}} & \textsc{\getVal{muL6}} & \textsc{\getVal{muL8}} & \textsc{\getVal{muG1}} & \multicolumn{1}{c}{\textsc{\getVal{muG2}}} \\ 
\midrule
\nopagebreak A & \nopagebreak 4.1.1  & \multicolumn{1}{r}{$\phantom{0}10$} & \multicolumn{1}{r}{$\phantom{0}2$} & \multicolumn{1}{r}{$\phantom{0}20${\tiny $\,\%$}} & \multicolumn{1}{r}{$\phantom{00}9${\tiny $\,\%$}} & \multicolumn{1}{r}{$\phantom{00}5${\tiny $\,\%$}} & \multicolumn{1}{r}{$\phantom{0}45${\tiny $\,\%$}} & \multicolumn{1}{r}{ 1} & \multicolumn{1}{r}{ 0} & \multicolumn{1}{r}{ 1} & \multicolumn{1}{r}{ 0} & \multicolumn{1}{r}{ 0} & \multicolumn{1}{r}{ 0} & \multicolumn{1}{r}{ 1} & \multicolumn{1}{r}{ 1} & \multicolumn{1}{r}{ 0} & \multicolumn{1}{r}{ 0} & \multicolumn{1}{r}{ 0} \\
 & \nopagebreak 4.3.2  & \multicolumn{1}{r}{$\phantom{0}10$} & \multicolumn{1}{r}{$\phantom{0}1$} & \multicolumn{1}{r}{$\phantom{0}10${\tiny $\,\%$}} & \multicolumn{1}{r}{$\phantom{00}4${\tiny $\,\%$}} & \multicolumn{1}{r}{$\phantom{00}0${\tiny $\,\%$}} & \multicolumn{1}{r}{$\phantom{0}42${\tiny $\,\%$}} & \multicolumn{1}{r}{ 0} & \multicolumn{1}{r}{ 0} & \multicolumn{1}{r}{ 0} & \multicolumn{1}{r}{ 0} & \multicolumn{1}{r}{ 0} & \multicolumn{1}{r}{ 0} & \multicolumn{1}{r}{ 1} & \multicolumn{1}{r}{ 1} & \multicolumn{1}{r}{ 0} & \multicolumn{1}{r}{ 0} & \multicolumn{1}{r}{ 0} \\
 & \nopagebreak 4.4.1  & \multicolumn{1}{r}{$\phantom{0}10$} & \multicolumn{1}{r}{$\phantom{0}1$} & \multicolumn{1}{r}{$\phantom{0}10${\tiny $\,\%$}} & \multicolumn{1}{r}{$\phantom{00}4${\tiny $\,\%$}} & \multicolumn{1}{r}{$\phantom{00}0${\tiny $\,\%$}} & \multicolumn{1}{r}{$\phantom{0}42${\tiny $\,\%$}} & \multicolumn{1}{r}{ 0} & \multicolumn{1}{r}{ 0} & \multicolumn{1}{r}{ 0} & \multicolumn{1}{r}{ 0} & \multicolumn{1}{r}{ 0} & \multicolumn{1}{r}{ 0} & \multicolumn{1}{r}{ 1} & \multicolumn{1}{r}{ 1} & \multicolumn{1}{r}{ 0} & \multicolumn{1}{r}{ 0} & \multicolumn{1}{r}{ 0} \\
 & \nopagebreak 4.5.0  & \multicolumn{1}{r}{$\phantom{0}10$} & \multicolumn{1}{r}{$\phantom{0}1$} & \multicolumn{1}{r}{$\phantom{0}10${\tiny $\,\%$}} & \multicolumn{1}{r}{$\phantom{00}4${\tiny $\,\%$}} & \multicolumn{1}{r}{$\phantom{00}0${\tiny $\,\%$}} & \multicolumn{1}{r}{$\phantom{0}42${\tiny $\,\%$}} & \multicolumn{1}{r}{ 0} & \multicolumn{1}{r}{ 0} & \multicolumn{1}{r}{ 0} & \multicolumn{1}{r}{ 0} & \multicolumn{1}{r}{ 0} & \multicolumn{1}{r}{ 0} & \multicolumn{1}{r}{ 1} & \multicolumn{1}{r}{ 1} & \multicolumn{1}{r}{ 0} & \multicolumn{1}{r}{ 0} & \multicolumn{1}{r}{ 0} \\
 & \nopagebreak \textsc{cvc4}  & \multicolumn{1}{r}{$\phantom{00}6$} & \multicolumn{1}{r}{$\phantom{0}0$} & \multicolumn{1}{r}{$\phantom{00}0${\tiny $\,\%$}} & \multicolumn{1}{r}{$\phantom{00}0${\tiny $\,\%$}} & \multicolumn{1}{r}{$\phantom{00}0${\tiny $\,\%$}} & \multicolumn{1}{r}{--} & \multicolumn{1}{r}{ 0} & \multicolumn{1}{r}{ 0} & \multicolumn{1}{r}{ 0} & \multicolumn{1}{r}{ 0} & \multicolumn{1}{r}{ 0} & \multicolumn{1}{r}{ 0} & \multicolumn{1}{r}{ 0} & \multicolumn{1}{r}{ 0} & \multicolumn{1}{r}{ 0} & \multicolumn{1}{r}{ 0} & \multicolumn{1}{r}{ 0} \\
 & \nopagebreak \textsc{why3}  & \multicolumn{1}{r}{$\phantom{00}7$} & \multicolumn{1}{r}{$\phantom{0}0$} & \multicolumn{1}{r}{$\phantom{00}0${\tiny $\,\%$}} & \multicolumn{1}{r}{$\phantom{00}0${\tiny $\,\%$}} & \multicolumn{1}{r}{$\phantom{00}0${\tiny $\,\%$}} & \multicolumn{1}{r}{--} & \multicolumn{1}{r}{} & \multicolumn{1}{r}{} & \multicolumn{1}{r}{} & \multicolumn{1}{r}{} & \multicolumn{1}{r}{} & \multicolumn{1}{r}{} & \multicolumn{1}{r}{} & \multicolumn{1}{r}{} & \multicolumn{1}{r}{} & \multicolumn{1}{r}{} & \multicolumn{1}{r}{} \\
\rule{0pt}{0.9\normalbaselineskip}D & \nopagebreak 4.1.1  & \multicolumn{1}{r}{$\phantom{00}0$} & \multicolumn{1}{r}{$\phantom{0}0$} & \multicolumn{1}{r}{--} & \multicolumn{1}{r}{--} & \multicolumn{1}{r}{--} & \multicolumn{1}{r}{--} & \multicolumn{1}{r}{ 0} & \multicolumn{1}{r}{ 0} & \multicolumn{1}{r}{ 0} & \multicolumn{1}{r}{ 0} & \multicolumn{1}{r}{ 0} & \multicolumn{1}{r}{ 0} & \multicolumn{1}{r}{ 0} & \multicolumn{1}{r}{ 0} & \multicolumn{1}{r}{ 0} & \multicolumn{1}{r}{ 0} & \multicolumn{1}{r}{ 0} \\
 & \nopagebreak 4.3.2  & \multicolumn{1}{r}{$\phantom{0}24$} & \multicolumn{1}{r}{$\phantom{0}7$} & \multicolumn{1}{r}{$\phantom{0}29${\tiny $\,\%$}} & \multicolumn{1}{r}{$\phantom{00}7${\tiny $\,\%$}} & \multicolumn{1}{r}{$\phantom{00}4${\tiny $\,\%$}} & \multicolumn{1}{r}{$\phantom{0}23${\tiny $\,\%$}} & \multicolumn{1}{r}{ 6} & \multicolumn{1}{r}{ 0} & \multicolumn{1}{r}{ 5} & \multicolumn{1}{r}{ 0} & \multicolumn{1}{r}{ 0} & \multicolumn{1}{r}{ 5} & \multicolumn{1}{r}{ 4} & \multicolumn{1}{r}{ 4} & \multicolumn{1}{r}{ 0} & \multicolumn{1}{r}{ 3} & \multicolumn{1}{r}{17} \\
 & \nopagebreak 4.4.1  & \multicolumn{1}{r}{$\phantom{0}24$} & \multicolumn{1}{r}{$\phantom{0}7$} & \multicolumn{1}{r}{$\phantom{0}29${\tiny $\,\%$}} & \multicolumn{1}{r}{$\phantom{00}7${\tiny $\,\%$}} & \multicolumn{1}{r}{$\phantom{00}5${\tiny $\,\%$}} & \multicolumn{1}{r}{$\phantom{0}23${\tiny $\,\%$}} & \multicolumn{1}{r}{ 6} & \multicolumn{1}{r}{ 0} & \multicolumn{1}{r}{ 5} & \multicolumn{1}{r}{ 0} & \multicolumn{1}{r}{ 0} & \multicolumn{1}{r}{ 5} & \multicolumn{1}{r}{ 4} & \multicolumn{1}{r}{ 4} & \multicolumn{1}{r}{ 0} & \multicolumn{1}{r}{ 3} & \multicolumn{1}{r}{17} \\
 & \nopagebreak 4.5.0  & \multicolumn{1}{r}{$\phantom{0}24$} & \multicolumn{1}{r}{$\phantom{0}6$} & \multicolumn{1}{r}{$\phantom{0}25${\tiny $\,\%$}} & \multicolumn{1}{r}{$\phantom{00}8${\tiny $\,\%$}} & \multicolumn{1}{r}{$\phantom{00}6${\tiny $\,\%$}} & \multicolumn{1}{r}{$\phantom{0}33${\tiny $\,\%$}} & \multicolumn{1}{r}{ 5} & \multicolumn{1}{r}{ 0} & \multicolumn{1}{r}{ 5} & \multicolumn{1}{r}{ 0} & \multicolumn{1}{r}{ 0} & \multicolumn{1}{r}{ 8} & \multicolumn{1}{r}{ 4} & \multicolumn{1}{r}{ 4} & \multicolumn{1}{r}{ 0} & \multicolumn{1}{r}{ 1} & \multicolumn{1}{r}{17} \\
 & \nopagebreak \textsc{cvc4}  & \multicolumn{1}{r}{$\phantom{00}0$} & \multicolumn{1}{r}{$\phantom{0}0$} & \multicolumn{1}{r}{--} & \multicolumn{1}{r}{--} & \multicolumn{1}{r}{--} & \multicolumn{1}{r}{--} & \multicolumn{1}{r}{ 0} & \multicolumn{1}{r}{ 0} & \multicolumn{1}{r}{ 0} & \multicolumn{1}{r}{ 0} & \multicolumn{1}{r}{ 0} & \multicolumn{1}{r}{ 0} & \multicolumn{1}{r}{ 0} & \multicolumn{1}{r}{ 0} & \multicolumn{1}{r}{ 0} & \multicolumn{1}{r}{ 0} & \multicolumn{1}{r}{ 0} \\
 & \nopagebreak \textsc{why3}  & \multicolumn{1}{r}{$\phantom{00}0$} & \multicolumn{1}{r}{$\phantom{0}0$} & \multicolumn{1}{r}{--} & \multicolumn{1}{r}{--} & \multicolumn{1}{r}{--} & \multicolumn{1}{r}{--} & \multicolumn{1}{r}{} & \multicolumn{1}{r}{} & \multicolumn{1}{r}{} & \multicolumn{1}{r}{} & \multicolumn{1}{r}{} & \multicolumn{1}{r}{} & \multicolumn{1}{r}{} & \multicolumn{1}{r}{} & \multicolumn{1}{r}{} & \multicolumn{1}{r}{} & \multicolumn{1}{r}{} \\
\rule{0pt}{0.9\normalbaselineskip}E & \nopagebreak 4.1.1  & \multicolumn{1}{r}{$\phantom{00}5$} & \multicolumn{1}{r}{$\phantom{0}0$} & \multicolumn{1}{r}{$\phantom{00}0${\tiny $\,\%$}} & \multicolumn{1}{r}{$\phantom{00}0${\tiny $\,\%$}} & \multicolumn{1}{r}{$\phantom{00}0${\tiny $\,\%$}} & \multicolumn{1}{r}{--} & \multicolumn{1}{r}{ 0} & \multicolumn{1}{r}{ 0} & \multicolumn{1}{r}{ 0} & \multicolumn{1}{r}{ 0} & \multicolumn{1}{r}{ 0} & \multicolumn{1}{r}{ 0} & \multicolumn{1}{r}{ 0} & \multicolumn{1}{r}{ 0} & \multicolumn{1}{r}{ 0} & \multicolumn{1}{r}{ 0} & \multicolumn{1}{r}{ 3} \\
 & \nopagebreak 4.3.2  & \multicolumn{1}{r}{$\phantom{00}7$} & \multicolumn{1}{r}{$\phantom{0}0$} & \multicolumn{1}{r}{$\phantom{00}0${\tiny $\,\%$}} & \multicolumn{1}{r}{$\phantom{00}0${\tiny $\,\%$}} & \multicolumn{1}{r}{$\phantom{00}0${\tiny $\,\%$}} & \multicolumn{1}{r}{--} & \multicolumn{1}{r}{ 0} & \multicolumn{1}{r}{ 0} & \multicolumn{1}{r}{ 0} & \multicolumn{1}{r}{ 0} & \multicolumn{1}{r}{ 0} & \multicolumn{1}{r}{ 0} & \multicolumn{1}{r}{ 0} & \multicolumn{1}{r}{ 0} & \multicolumn{1}{r}{ 0} & \multicolumn{1}{r}{ 0} & \multicolumn{1}{r}{ 3} \\
 & \nopagebreak 4.4.1  & \multicolumn{1}{r}{$\phantom{00}7$} & \multicolumn{1}{r}{$\phantom{0}0$} & \multicolumn{1}{r}{$\phantom{00}0${\tiny $\,\%$}} & \multicolumn{1}{r}{$\phantom{00}0${\tiny $\,\%$}} & \multicolumn{1}{r}{$\phantom{00}0${\tiny $\,\%$}} & \multicolumn{1}{r}{--} & \multicolumn{1}{r}{ 0} & \multicolumn{1}{r}{ 0} & \multicolumn{1}{r}{ 0} & \multicolumn{1}{r}{ 0} & \multicolumn{1}{r}{ 0} & \multicolumn{1}{r}{ 0} & \multicolumn{1}{r}{ 0} & \multicolumn{1}{r}{ 0} & \multicolumn{1}{r}{ 0} & \multicolumn{1}{r}{ 0} & \multicolumn{1}{r}{ 3} \\
 & \nopagebreak 4.5.0  & \multicolumn{1}{r}{$\phantom{00}7$} & \multicolumn{1}{r}{$\phantom{0}0$} & \multicolumn{1}{r}{$\phantom{00}0${\tiny $\,\%$}} & \multicolumn{1}{r}{$\phantom{00}0${\tiny $\,\%$}} & \multicolumn{1}{r}{$\phantom{00}0${\tiny $\,\%$}} & \multicolumn{1}{r}{--} & \multicolumn{1}{r}{ 0} & \multicolumn{1}{r}{ 0} & \multicolumn{1}{r}{ 0} & \multicolumn{1}{r}{ 0} & \multicolumn{1}{r}{ 0} & \multicolumn{1}{r}{ 0} & \multicolumn{1}{r}{ 0} & \multicolumn{1}{r}{ 0} & \multicolumn{1}{r}{ 0} & \multicolumn{1}{r}{ 0} & \multicolumn{1}{r}{ 3} \\
 & \nopagebreak \textsc{cvc4}  & \multicolumn{1}{r}{$\phantom{00}4$} & \multicolumn{1}{r}{$\phantom{0}0$} & \multicolumn{1}{r}{$\phantom{00}0${\tiny $\,\%$}} & \multicolumn{1}{r}{$\phantom{00}0${\tiny $\,\%$}} & \multicolumn{1}{r}{$\phantom{00}0${\tiny $\,\%$}} & \multicolumn{1}{r}{--} & \multicolumn{1}{r}{ 0} & \multicolumn{1}{r}{ 0} & \multicolumn{1}{r}{ 0} & \multicolumn{1}{r}{ 0} & \multicolumn{1}{r}{ 0} & \multicolumn{1}{r}{ 0} & \multicolumn{1}{r}{ 0} & \multicolumn{1}{r}{ 0} & \multicolumn{1}{r}{ 0} & \multicolumn{1}{r}{ 0} & \multicolumn{1}{r}{ 3} \\
 & \nopagebreak \textsc{why3}  & \multicolumn{1}{r}{$\phantom{00}7$} & \multicolumn{1}{r}{$\phantom{0}0$} & \multicolumn{1}{r}{$\phantom{00}0${\tiny $\,\%$}} & \multicolumn{1}{r}{$\phantom{00}0${\tiny $\,\%$}} & \multicolumn{1}{r}{$\phantom{00}0${\tiny $\,\%$}} & \multicolumn{1}{r}{--} & \multicolumn{1}{r}{} & \multicolumn{1}{r}{} & \multicolumn{1}{r}{} & \multicolumn{1}{r}{} & \multicolumn{1}{r}{} & \multicolumn{1}{r}{} & \multicolumn{1}{r}{} & \multicolumn{1}{r}{} & \multicolumn{1}{r}{} & \multicolumn{1}{r}{} & \multicolumn{1}{r}{} \\
\rule{0pt}{0.9\normalbaselineskip}P & \nopagebreak 4.1.1  & \multicolumn{1}{r}{$\phantom{00}0$} & \multicolumn{1}{r}{$\phantom{0}0$} & \multicolumn{1}{r}{--} & \multicolumn{1}{r}{--} & \multicolumn{1}{r}{--} & \multicolumn{1}{r}{--} & \multicolumn{1}{r}{ 0} & \multicolumn{1}{r}{ 0} & \multicolumn{1}{r}{ 0} & \multicolumn{1}{r}{ 0} & \multicolumn{1}{r}{ 0} & \multicolumn{1}{r}{ 0} & \multicolumn{1}{r}{ 0} & \multicolumn{1}{r}{ 0} & \multicolumn{1}{r}{ 0} & \multicolumn{1}{r}{ 0} & \multicolumn{1}{r}{ 0} \\
 & \nopagebreak 4.3.2  & \multicolumn{1}{r}{$\phantom{0}14$} & \multicolumn{1}{r}{$\phantom{0}6$} & \multicolumn{1}{r}{$\phantom{0}43${\tiny $\,\%$}} & \multicolumn{1}{r}{$\phantom{00}1${\tiny $\,\%$}} & \multicolumn{1}{r}{$\phantom{00}0${\tiny $\,\%$}} & \multicolumn{1}{r}{$\phantom{00}2${\tiny $\,\%$}} & \multicolumn{1}{r}{ 4} & \multicolumn{1}{r}{ 0} & \multicolumn{1}{r}{ 7} & \multicolumn{1}{r}{ 0} & \multicolumn{1}{r}{ 0} & \multicolumn{1}{r}{ 1} & \multicolumn{1}{r}{ 1} & \multicolumn{1}{r}{ 0} & \multicolumn{1}{r}{ 0} & \multicolumn{1}{r}{ 0} & \multicolumn{1}{r}{10} \\
 & \nopagebreak 4.4.1  & \multicolumn{1}{r}{$\phantom{0}13$} & \multicolumn{1}{r}{$\phantom{0}5$} & \multicolumn{1}{r}{$\phantom{0}38${\tiny $\,\%$}} & \multicolumn{1}{r}{$\phantom{00}1${\tiny $\,\%$}} & \multicolumn{1}{r}{$\phantom{00}0${\tiny $\,\%$}} & \multicolumn{1}{r}{$\phantom{00}2${\tiny $\,\%$}} & \multicolumn{1}{r}{ 3} & \multicolumn{1}{r}{ 0} & \multicolumn{1}{r}{ 6} & \multicolumn{1}{r}{ 0} & \multicolumn{1}{r}{ 0} & \multicolumn{1}{r}{ 0} & \multicolumn{1}{r}{ 0} & \multicolumn{1}{r}{ 0} & \multicolumn{1}{r}{ 0} & \multicolumn{1}{r}{ 0} & \multicolumn{1}{r}{ 9} \\
 & \nopagebreak 4.5.0  & \multicolumn{1}{r}{$\phantom{0}13$} & \multicolumn{1}{r}{$\phantom{0}5$} & \multicolumn{1}{r}{$\phantom{0}38${\tiny $\,\%$}} & \multicolumn{1}{r}{$\phantom{00}1${\tiny $\,\%$}} & \multicolumn{1}{r}{$\phantom{00}0${\tiny $\,\%$}} & \multicolumn{1}{r}{$\phantom{00}2${\tiny $\,\%$}} & \multicolumn{1}{r}{ 4} & \multicolumn{1}{r}{ 0} & \multicolumn{1}{r}{ 6} & \multicolumn{1}{r}{ 0} & \multicolumn{1}{r}{ 0} & \multicolumn{1}{r}{ 0} & \multicolumn{1}{r}{ 0} & \multicolumn{1}{r}{ 0} & \multicolumn{1}{r}{ 0} & \multicolumn{1}{r}{ 0} & \multicolumn{1}{r}{ 9} \\
 & \nopagebreak \textsc{cvc4}  & \multicolumn{1}{r}{$\phantom{00}0$} & \multicolumn{1}{r}{$\phantom{0}0$} & \multicolumn{1}{r}{--} & \multicolumn{1}{r}{--} & \multicolumn{1}{r}{--} & \multicolumn{1}{r}{--} & \multicolumn{1}{r}{ 0} & \multicolumn{1}{r}{ 0} & \multicolumn{1}{r}{ 0} & \multicolumn{1}{r}{ 0} & \multicolumn{1}{r}{ 0} & \multicolumn{1}{r}{ 0} & \multicolumn{1}{r}{ 0} & \multicolumn{1}{r}{ 0} & \multicolumn{1}{r}{ 0} & \multicolumn{1}{r}{ 0} & \multicolumn{1}{r}{ 0} \\
 & \nopagebreak \textsc{why3}  & \multicolumn{1}{r}{$\phantom{00}0$} & \multicolumn{1}{r}{$\phantom{0}0$} & \multicolumn{1}{r}{--} & \multicolumn{1}{r}{--} & \multicolumn{1}{r}{--} & \multicolumn{1}{r}{--} & \multicolumn{1}{r}{} & \multicolumn{1}{r}{} & \multicolumn{1}{r}{} & \multicolumn{1}{r}{} & \multicolumn{1}{r}{} & \multicolumn{1}{r}{} & \multicolumn{1}{r}{} & \multicolumn{1}{r}{} & \multicolumn{1}{r}{} & \multicolumn{1}{r}{} & \multicolumn{1}{r}{} \\
\rule{0pt}{0.9\normalbaselineskip}S & \nopagebreak 4.1.1  & \multicolumn{1}{r}{$\phantom{0}51$} & \multicolumn{1}{r}{$\phantom{0}0$} & \multicolumn{1}{r}{$\phantom{00}0${\tiny $\,\%$}} & \multicolumn{1}{r}{$\phantom{00}0${\tiny $\,\%$}} & \multicolumn{1}{r}{$\phantom{00}0${\tiny $\,\%$}} & \multicolumn{1}{r}{--} & \multicolumn{1}{r}{ 0} & \multicolumn{1}{r}{ 0} & \multicolumn{1}{r}{ 0} & \multicolumn{1}{r}{ 0} & \multicolumn{1}{r}{ 0} & \multicolumn{1}{r}{ 0} & \multicolumn{1}{r}{ 0} & \multicolumn{1}{r}{ 0} & \multicolumn{1}{r}{ 0} & \multicolumn{1}{r}{ 0} & \multicolumn{1}{r}{ 0} \\
 & \nopagebreak 4.3.2  & \multicolumn{1}{r}{$\phantom{0}51$} & \multicolumn{1}{r}{$\phantom{0}0$} & \multicolumn{1}{r}{$\phantom{00}0${\tiny $\,\%$}} & \multicolumn{1}{r}{$\phantom{00}0${\tiny $\,\%$}} & \multicolumn{1}{r}{$\phantom{00}0${\tiny $\,\%$}} & \multicolumn{1}{r}{--} & \multicolumn{1}{r}{ 0} & \multicolumn{1}{r}{ 0} & \multicolumn{1}{r}{ 0} & \multicolumn{1}{r}{ 0} & \multicolumn{1}{r}{ 0} & \multicolumn{1}{r}{ 0} & \multicolumn{1}{r}{ 0} & \multicolumn{1}{r}{ 0} & \multicolumn{1}{r}{ 0} & \multicolumn{1}{r}{ 0} & \multicolumn{1}{r}{ 0} \\
 & \nopagebreak 4.4.1  & \multicolumn{1}{r}{$\phantom{0}51$} & \multicolumn{1}{r}{$\phantom{0}0$} & \multicolumn{1}{r}{$\phantom{00}0${\tiny $\,\%$}} & \multicolumn{1}{r}{$\phantom{00}0${\tiny $\,\%$}} & \multicolumn{1}{r}{$\phantom{00}0${\tiny $\,\%$}} & \multicolumn{1}{r}{--} & \multicolumn{1}{r}{ 0} & \multicolumn{1}{r}{ 0} & \multicolumn{1}{r}{ 0} & \multicolumn{1}{r}{ 0} & \multicolumn{1}{r}{ 0} & \multicolumn{1}{r}{ 0} & \multicolumn{1}{r}{ 0} & \multicolumn{1}{r}{ 0} & \multicolumn{1}{r}{ 0} & \multicolumn{1}{r}{ 0} & \multicolumn{1}{r}{ 0} \\
 & \nopagebreak 4.5.0  & \multicolumn{1}{r}{$\phantom{0}51$} & \multicolumn{1}{r}{$\phantom{0}0$} & \multicolumn{1}{r}{$\phantom{00}0${\tiny $\,\%$}} & \multicolumn{1}{r}{$\phantom{00}0${\tiny $\,\%$}} & \multicolumn{1}{r}{$\phantom{00}0${\tiny $\,\%$}} & \multicolumn{1}{r}{--} & \multicolumn{1}{r}{ 0} & \multicolumn{1}{r}{ 0} & \multicolumn{1}{r}{ 0} & \multicolumn{1}{r}{ 0} & \multicolumn{1}{r}{ 0} & \multicolumn{1}{r}{ 0} & \multicolumn{1}{r}{ 0} & \multicolumn{1}{r}{ 0} & \multicolumn{1}{r}{ 0} & \multicolumn{1}{r}{ 0} & \multicolumn{1}{r}{ 0} \\
 & \nopagebreak \textsc{cvc4}  & \multicolumn{1}{r}{$\phantom{0}51$} & \multicolumn{1}{r}{$\phantom{0}0$} & \multicolumn{1}{r}{$\phantom{00}0${\tiny $\,\%$}} & \multicolumn{1}{r}{$\phantom{00}0${\tiny $\,\%$}} & \multicolumn{1}{r}{$\phantom{00}0${\tiny $\,\%$}} & \multicolumn{1}{r}{--} & \multicolumn{1}{r}{ 0} & \multicolumn{1}{r}{ 0} & \multicolumn{1}{r}{ 0} & \multicolumn{1}{r}{ 0} & \multicolumn{1}{r}{ 0} & \multicolumn{1}{r}{ 0} & \multicolumn{1}{r}{ 0} & \multicolumn{1}{r}{ 0} & \multicolumn{1}{r}{ 0} & \multicolumn{1}{r}{ 0} & \multicolumn{1}{r}{ 0} \\
 & \nopagebreak \textsc{why3}  & \multicolumn{1}{r}{$\phantom{0}40$} & \multicolumn{1}{r}{$\phantom{0}2$} & \multicolumn{1}{r}{$\phantom{00}5${\tiny $\,\%$}} & \multicolumn{1}{r}{$\phantom{00}1${\tiny $\,\%$}} & \multicolumn{1}{r}{$\phantom{00}1${\tiny $\,\%$}} & \multicolumn{1}{r}{$\phantom{0}25${\tiny $\,\%$}} & \multicolumn{1}{r}{} & \multicolumn{1}{r}{} & \multicolumn{1}{r}{} & \multicolumn{1}{r}{} & \multicolumn{1}{r}{} & \multicolumn{1}{r}{} & \multicolumn{1}{r}{} & \multicolumn{1}{r}{} & \multicolumn{1}{r}{} & \multicolumn{1}{r}{} & \multicolumn{1}{r}{} \\
\rule{0pt}{0.9\normalbaselineskip}T & \nopagebreak 4.1.1  & \multicolumn{1}{r}{$\phantom{00}8$} & \multicolumn{1}{r}{$\phantom{0}1$} & \multicolumn{1}{r}{$\phantom{0}12${\tiny $\,\%$}} & \multicolumn{1}{r}{$\phantom{00}5${\tiny $\,\%$}} & \multicolumn{1}{r}{$\phantom{00}5${\tiny $\,\%$}} & \multicolumn{1}{r}{$\phantom{0}39${\tiny $\,\%$}} & \multicolumn{1}{r}{ 1} & \multicolumn{1}{r}{ 0} & \multicolumn{1}{r}{ 1} & \multicolumn{1}{r}{ 0} & \multicolumn{1}{r}{ 0} & \multicolumn{1}{r}{ 1} & \multicolumn{1}{r}{ 1} & \multicolumn{1}{r}{ 1} & \multicolumn{1}{r}{ 0} & \multicolumn{1}{r}{ 1} & \multicolumn{1}{r}{ 1} \\
 & \nopagebreak 4.3.2  & \multicolumn{1}{r}{$\phantom{00}8$} & \multicolumn{1}{r}{$\phantom{0}1$} & \multicolumn{1}{r}{$\phantom{0}12${\tiny $\,\%$}} & \multicolumn{1}{r}{$\phantom{00}8${\tiny $\,\%$}} & \multicolumn{1}{r}{$\phantom{00}8${\tiny $\,\%$}} & \multicolumn{1}{r}{$\phantom{0}62${\tiny $\,\%$}} & \multicolumn{1}{r}{ 1} & \multicolumn{1}{r}{ 0} & \multicolumn{1}{r}{ 1} & \multicolumn{1}{r}{ 0} & \multicolumn{1}{r}{ 0} & \multicolumn{1}{r}{ 1} & \multicolumn{1}{r}{ 1} & \multicolumn{1}{r}{ 1} & \multicolumn{1}{r}{ 0} & \multicolumn{1}{r}{ 1} & \multicolumn{1}{r}{ 1} \\
 & \nopagebreak 4.4.1  & \multicolumn{1}{r}{$\phantom{00}8$} & \multicolumn{1}{r}{$\phantom{0}1$} & \multicolumn{1}{r}{$\phantom{0}12${\tiny $\,\%$}} & \multicolumn{1}{r}{$\phantom{00}8${\tiny $\,\%$}} & \multicolumn{1}{r}{$\phantom{00}8${\tiny $\,\%$}} & \multicolumn{1}{r}{$\phantom{0}60${\tiny $\,\%$}} & \multicolumn{1}{r}{ 1} & \multicolumn{1}{r}{ 0} & \multicolumn{1}{r}{ 1} & \multicolumn{1}{r}{ 0} & \multicolumn{1}{r}{ 0} & \multicolumn{1}{r}{ 1} & \multicolumn{1}{r}{ 1} & \multicolumn{1}{r}{ 1} & \multicolumn{1}{r}{ 0} & \multicolumn{1}{r}{ 1} & \multicolumn{1}{r}{ 1} \\
 & \nopagebreak 4.5.0  & \multicolumn{1}{r}{$\phantom{00}8$} & \multicolumn{1}{r}{$\phantom{0}1$} & \multicolumn{1}{r}{$\phantom{0}12${\tiny $\,\%$}} & \multicolumn{1}{r}{$\phantom{0}12${\tiny $\,\%$}} & \multicolumn{1}{r}{$\phantom{0}12${\tiny $\,\%$}} & \multicolumn{1}{r}{$\phantom{0}96${\tiny $\,\%$}} & \multicolumn{1}{r}{ 1} & \multicolumn{1}{r}{ 0} & \multicolumn{1}{r}{ 1} & \multicolumn{1}{r}{ 0} & \multicolumn{1}{r}{ 0} & \multicolumn{1}{r}{ 1} & \multicolumn{1}{r}{ 1} & \multicolumn{1}{r}{ 1} & \multicolumn{1}{r}{ 0} & \multicolumn{1}{r}{ 1} & \multicolumn{1}{r}{ 1} \\
 & \nopagebreak \textsc{cvc4}  & \multicolumn{1}{r}{$\phantom{00}4$} & \multicolumn{1}{r}{$\phantom{0}0$} & \multicolumn{1}{r}{$\phantom{00}0${\tiny $\,\%$}} & \multicolumn{1}{r}{$\phantom{00}0${\tiny $\,\%$}} & \multicolumn{1}{r}{$\phantom{00}0${\tiny $\,\%$}} & \multicolumn{1}{r}{--} & \multicolumn{1}{r}{ 0} & \multicolumn{1}{r}{ 0} & \multicolumn{1}{r}{ 0} & \multicolumn{1}{r}{ 0} & \multicolumn{1}{r}{ 0} & \multicolumn{1}{r}{ 0} & \multicolumn{1}{r}{ 0} & \multicolumn{1}{r}{ 0} & \multicolumn{1}{r}{ 0} & \multicolumn{1}{r}{ 0} & \multicolumn{1}{r}{ 1} \\
 & \nopagebreak \textsc{why3}  & \multicolumn{1}{r}{$\phantom{00}3$} & \multicolumn{1}{r}{$\phantom{0}0$} & \multicolumn{1}{r}{$\phantom{00}0${\tiny $\,\%$}} & \multicolumn{1}{r}{$\phantom{00}0${\tiny $\,\%$}} & \multicolumn{1}{r}{$\phantom{00}0${\tiny $\,\%$}} & \multicolumn{1}{r}{--} & \multicolumn{1}{r}{} & \multicolumn{1}{r}{} & \multicolumn{1}{r}{} & \multicolumn{1}{r}{} & \multicolumn{1}{r}{} & \multicolumn{1}{r}{} & \multicolumn{1}{r}{} & \multicolumn{1}{r}{} & \multicolumn{1}{r}{} & \multicolumn{1}{r}{} & \multicolumn{1}{r}{} \\
\cmidrule(){2-19}
\rule{0pt}{0.9\normalbaselineskip}\textsl{all} & \nopagebreak 4.1.1  & \multicolumn{1}{r}{$\phantom{0}74$} & \multicolumn{1}{r}{$\phantom{0}3$} & \multicolumn{1}{r}{$\phantom{00}4${\tiny $\,\%$}} & \multicolumn{1}{r}{$\phantom{00}2${\tiny $\,\%$}} & \multicolumn{1}{r}{$\phantom{00}1${\tiny $\,\%$}} & \multicolumn{1}{r}{$\phantom{0}43${\tiny $\,\%$}} & \multicolumn{1}{r}{ 2} & \multicolumn{1}{r}{ 0} & \multicolumn{1}{r}{ 2} & \multicolumn{1}{r}{ 0} & \multicolumn{1}{r}{ 0} & \multicolumn{1}{r}{ 1} & \multicolumn{1}{r}{ 2} & \multicolumn{1}{r}{ 2} & \multicolumn{1}{r}{ 0} & \multicolumn{1}{r}{ 1} & \multicolumn{1}{r}{ 4} \\
 & \nopagebreak 4.3.2  & \multicolumn{1}{r}{$114$} & \multicolumn{1}{r}{$15$} & \multicolumn{1}{r}{$\phantom{0}13${\tiny $\,\%$}} & \multicolumn{1}{r}{$\phantom{00}2${\tiny $\,\%$}} & \multicolumn{1}{r}{$\phantom{00}2${\tiny $\,\%$}} & \multicolumn{1}{r}{$\phantom{0}18${\tiny $\,\%$}} & \multicolumn{1}{r}{11} & \multicolumn{1}{r}{ 0} & \multicolumn{1}{r}{13} & \multicolumn{1}{r}{ 0} & \multicolumn{1}{r}{ 0} & \multicolumn{1}{r}{ 7} & \multicolumn{1}{r}{ 7} & \multicolumn{1}{r}{ 6} & \multicolumn{1}{r}{ 0} & \multicolumn{1}{r}{ 4} & \multicolumn{1}{r}{31} \\
 & \nopagebreak 4.4.1  & \multicolumn{1}{r}{$113$} & \multicolumn{1}{r}{$14$} & \multicolumn{1}{r}{$\phantom{0}12${\tiny $\,\%$}} & \multicolumn{1}{r}{$\phantom{00}2${\tiny $\,\%$}} & \multicolumn{1}{r}{$\phantom{00}2${\tiny $\,\%$}} & \multicolumn{1}{r}{$\phantom{0}20${\tiny $\,\%$}} & \multicolumn{1}{r}{10} & \multicolumn{1}{r}{ 0} & \multicolumn{1}{r}{12} & \multicolumn{1}{r}{ 0} & \multicolumn{1}{r}{ 0} & \multicolumn{1}{r}{ 6} & \multicolumn{1}{r}{ 6} & \multicolumn{1}{r}{ 6} & \multicolumn{1}{r}{ 0} & \multicolumn{1}{r}{ 4} & \multicolumn{1}{r}{30} \\
 & \nopagebreak 4.5.0  & \multicolumn{1}{r}{$113$} & \multicolumn{1}{r}{$13$} & \multicolumn{1}{r}{$\phantom{0}12${\tiny $\,\%$}} & \multicolumn{1}{r}{$\phantom{00}3${\tiny $\,\%$}} & \multicolumn{1}{r}{$\phantom{00}2${\tiny $\,\%$}} & \multicolumn{1}{r}{$\phantom{0}27${\tiny $\,\%$}} & \multicolumn{1}{r}{10} & \multicolumn{1}{r}{ 0} & \multicolumn{1}{r}{12} & \multicolumn{1}{r}{ 0} & \multicolumn{1}{r}{ 0} & \multicolumn{1}{r}{ 9} & \multicolumn{1}{r}{ 6} & \multicolumn{1}{r}{ 6} & \multicolumn{1}{r}{ 0} & \multicolumn{1}{r}{ 2} & \multicolumn{1}{r}{30} \\
 & \nopagebreak \textsc{cvc4}  & \multicolumn{1}{r}{$\phantom{0}65$} & \multicolumn{1}{r}{$\phantom{0}0$} & \multicolumn{1}{r}{$\phantom{00}0${\tiny $\,\%$}} & \multicolumn{1}{r}{$\phantom{00}0${\tiny $\,\%$}} & \multicolumn{1}{r}{$\phantom{00}0${\tiny $\,\%$}} & \multicolumn{1}{r}{--} & \multicolumn{1}{r}{ 0} & \multicolumn{1}{r}{ 0} & \multicolumn{1}{r}{ 0} & \multicolumn{1}{r}{ 0} & \multicolumn{1}{r}{ 0} & \multicolumn{1}{r}{ 0} & \multicolumn{1}{r}{ 0} & \multicolumn{1}{r}{ 0} & \multicolumn{1}{r}{ 0} & \multicolumn{1}{r}{ 0} & \multicolumn{1}{r}{ 4} \\
 & \nopagebreak \textsc{why3}  & \multicolumn{1}{r}{$\phantom{0}57$} & \multicolumn{1}{r}{$\phantom{0}2$} & \multicolumn{1}{r}{$\phantom{00}4${\tiny $\,\%$}} & \multicolumn{1}{r}{$\phantom{00}1${\tiny $\,\%$}} & \multicolumn{1}{r}{$\phantom{00}1${\tiny $\,\%$}} & \multicolumn{1}{r}{$\phantom{0}25${\tiny $\,\%$}} & \multicolumn{1}{r}{} & \multicolumn{1}{r}{} & \multicolumn{1}{r}{} & \multicolumn{1}{r}{} & \multicolumn{1}{r}{} & \multicolumn{1}{r}{} & \multicolumn{1}{r}{} & \multicolumn{1}{r}{} & \multicolumn{1}{r}{} & \multicolumn{1}{r}{} & \multicolumn{1}{r}{} \\
\bottomrule 
\end{tabular}
  
\caption{
\scriptsize
{Experimental results of robustness testing with \mugie.}
{
For each \textsc{group} of seeds, for each \textsc{tool}:
number of seeds passing verification (\textsc{\# pass}),
number and percentage of passing seeds for which at least one mutant fails verification (\textsc{\# $\exists$fail} and \textsc{\% $\exists$fail}),
 average percentage of mutants per passing seed that fail verification ($\overline{\textsc{\% fail}}$),
average percentage of mutants per passing seed that time out ($\overline{\textsc{\% timeout}}$),
average percentage of mutants that fail verification per passing seed with at least one failing mutant ($\overline{\textsc{\% fail}}$).
The middle section of the table records experiments with batch $M_*$;
each of the 11 rightmost columns records experiments with batch $M_J$, for $J$ one of the mutation operators in \autoref{tab:transformations}.
}
}
\label{tab:fail}
\end{table}

\subsection{Experimental Results}
\label{sec:results}

\pparagraph{Overall results: batch $M_*$.}
Our experiments, whose detailed results are in \autoref{tab:fail},
show that
robustness testing is \emph{effective} in exposing \emph{brittle behavior}, 
which is \emph{recurrent} in Boogie:
for 12\% of the seeds that pass verification,\footnote{For clarity, we initially focus on Boogie~4.5.0, and later discuss differences with other versions.}
there is at least one mutant in batch $M_*$ that fails verification.

Not all seeds are equally prone to brittleness:
while on average only 3\% of one seed's mutants fail verification, 
it is considerably easier to trip up seeds that are \emph{susceptible} to brittle behavior 
(that is such that at least one mutant fails verification): 27\% of mutants per such seeds fail verification.

When the verifier times out on a mutant, it may be because:
\begin{enumerate*}[label=\emph{\roman*)}]
\item \label{to:spurious} the timeout is itself unstable and due to random noise in the runtime environment;
\item \label{to:longer} the mutant takes longer to verify than the seed, but may still be verified given longer time;
\item \label{to:infinity} verification time diverges.
\end{enumerate*}
We ruled out \ref{to:spurious} by repeating experiments 10 times, and reporting a timeout only if all 10 repetitions time out.
Thus, we can generally consider the timeouts in \autoref{tab:fail} indicative of a genuine degrading of performance in verification---which affected 3\% of one seed's mutants on average.

\pparagraph{Boogie versions.}
There is little difference between Boogie versions, with the exception of Boogie 4.1.1.
This older version does not support some language features used extensively in many larger examples that also tend to be more brittle (groups D and P).
As a result, the percentage of verified seeds with mutants that fail verification is spuriously lower (4\%) but only because the experiments with Boogie 4.1.1 dodged the harder problems and performed similarly to the other Boogie versions on the simpler ones.
\iflong
Leaving the older Boogie version 4.1.1 aside, 
our experiments leave open the question of whether robustness has significantly improved 
in recent versions of Boogie.
\fi

\pparagraph{Intermediate verifier vs.\ backend.}
Is the brittleness we observed in our experiments imputable to Boogie or really to Z3?
To shed light on this question, we tried to verify every seed and mutant using CVC4 instead of Z3 with Boogie's encoding; and using Why3 on a translation~\cite{b2w} of Boogie's input.
Since the seeds are programs optimized for Boogie verification, CVC4 and Why3 can correctly process only about half of the seeds that Boogie can.
This gives us too little evidence to answer the question conclusively: 
while both CVC4 and Why3 seem to be more robust than Boogie, 
they can verify none of the brittle seeds (that is, verified seeds with at least one failing mutant), and thus behave as robustly as Boogie on the programs that both tools can process.\footnote{Additionally, Why3 times out on 51 mutants of 2 seeds in group S; this seems to reflect an ineffective translation performed by \btw~\cite{b2w} rather than brittleness of Why3.}
As suggested by the simple example of \autoref{sec:example} (where Why3 was indeed more robust than Boogie), it is really the interplay of Boogie and Z3 that determines brittle behavior.
While SMT solvers have their own quirks, Boogie is meant to provide a stable intermediate layer;
in all, it seems fair to say that Boogie is at least partly responsible for the brittleness.

\pparagraph{Program groups.}
Robustness varies greatly across groups, according to features and complexity of the seeds that are mutated.
Groups D and P are the most brittle: about 1/3 of passing seeds in D, and about 2/5 of passing seeds in P, have at least one mutant that fails verification.
Seeds in D and P are large and complex programs generated by Dafny and AutoProof;
they include extensive definitions with plenty of generic types, complex axioms, and instantiations.
The brittleness of these programs reflects the hardness of verifying strong specifications and feature-rich programming languages: the Boogie encoding must be optimized in every aspect if it has to be automatically verifiable; even a modicum of clutter---introduced by \mugie---may jeopardize successful verification.

By the same token, groups A, E, and T's programs are more robust because they have a smaller impact surface in terms of features and size.
Group S's programs are uniformly robust 
because they have simple, repetitive structure and weak specifications despite their significant size.

\begin{figure}[!hbt]
\centering
\input{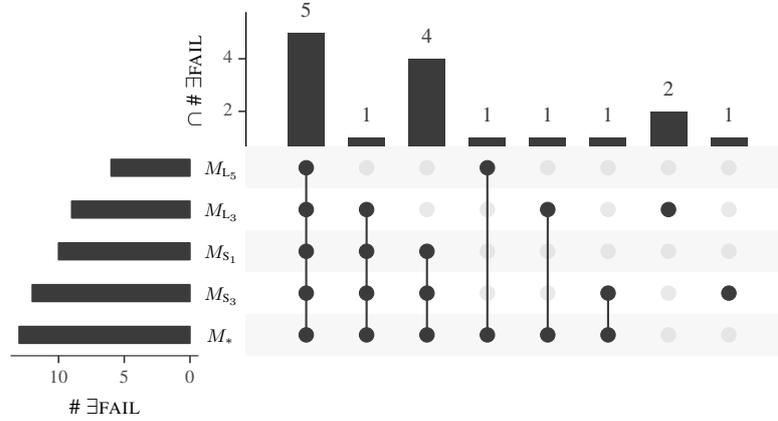}
\caption{\footnotesize For each of 16 verified seeds with at least one failing mutant with Boogie~4.5.0, which batches all exclusively include a failing mutant of those seeds.
\ref{m:G2} is excluded and analyzed separately;
\ref{m:S5}, \ref{m:L1}, \ref{m:L2}, \ref{m:L8} could not generate any failing mutant;
\ref{m:L5} generated failing mutants for a strict subset of those in $M_*$; 
\ref{m:G1} generated failing mutants for a strict subset of those in $M_\text{\ref{m:L6}}$.}
\label{fig:intersections}
\end{figure}

\pparagraph{Mutation operators and batches $M_J$.}
\autoref{fig:intersections} and the rightmost columns of \autoref{tab:fail}
explore the relative effectiveness of each mutation operator.
\getVal{muS5}, \getVal{muL1}, \getVal{muL2}, and \getVal{muL8} could not generate any failing mutant---suggesting that Boogie's encoding of procedure declarations, of local variables, and of conditionals is fairly robust.
In contrast, all other operators could generate at least one failing mutant;  \autoref{fig:intersections} indicates that \getVal{muL4} and \getVal{muS6} generated failing mutants for respectively 2 seeds and 1 seed that were robust in batch $M_*$ (using all mutation operators with the same frequency)---indicating that mutation operators are complementary to a certain extent in the kind of brittleness they can expose.
\iflong
Since every seed that $M_*$ trips up can also be tripped up by a single operator, 
combining multiple mutation operators does not seem to be necessary for successful robustness testing (although predicting which operators will be effective may be hard).
\fi

\iflong
\autoref{tab:muops} gives 
information about the frequency of different mutation operators in failing mutants of batch $M_*$.
Differences are consistent with the results of batches $M_J$---which provide more direct evidence.
As usual, the numbers for Boogie 4.1.1 are different because of a much smaller number of passing seeds on which the statistics are computed, which in turn is a result of Boogie 4.1.1's more limited supported features.

\begin{table}[!htb]
\centering
\scriptsize
\begin{tabular}{llcccccccccc}
\toprule
& &
\multicolumn{10}{c}{\textsc{\% failing mutants in batch $M_*$ per mutation operator}}
\\
\cmidrule(lr){3-12}
\textsc{group} & \textsc{tool} & \getVal{muL1} & \getVal{muL2} & \getVal{muL4} & \getVal{muL5} & \getVal{muL6} & \getVal{muL8} & \getVal{muS1} & \getVal{muS5} & \getVal{muS6} & \multicolumn{1}{c}{\getVal{muG1}} \\ 
\midrule
\nopagebreak A & \nopagebreak 4.1.1  & \multicolumn{1}{r}{$\phantom{00}7${\tiny $\,\%$}} & \multicolumn{1}{r}{$\phantom{0}43${\tiny $\,\%$}} & \multicolumn{1}{r}{$\phantom{0}36${\tiny $\,\%$}} & \multicolumn{1}{r}{$\phantom{0}57${\tiny $\,\%$}} & \multicolumn{1}{r}{$\phantom{0}34${\tiny $\,\%$}} & \multicolumn{1}{r}{$\phantom{00}6${\tiny $\,\%$}} & \multicolumn{1}{r}{$\phantom{0}91${\tiny $\,\%$}} & \multicolumn{1}{r}{$\phantom{0}33${\tiny $\,\%$}} & \multicolumn{1}{r}{$\phantom{0}18${\tiny $\,\%$}} & \multicolumn{1}{r}{$\phantom{0}38${\tiny $\,\%$}} \\
 & \nopagebreak 4.3.2  & \multicolumn{1}{r}{$\phantom{0}14${\tiny $\,\%$}} & \multicolumn{1}{r}{$\phantom{0}86${\tiny $\,\%$}} & \multicolumn{1}{r}{$\phantom{0}43${\tiny $\,\%$}} & \multicolumn{1}{r}{$\phantom{0}69${\tiny $\,\%$}} & \multicolumn{1}{r}{$\phantom{0}48${\tiny $\,\%$}} & \multicolumn{1}{r}{$\phantom{0}12${\tiny $\,\%$}} & \multicolumn{1}{r}{$\phantom{0}90${\tiny $\,\%$}} & \multicolumn{1}{r}{$\phantom{0}12${\tiny $\,\%$}} & \multicolumn{1}{r}{$\phantom{0}14${\tiny $\,\%$}} & \multicolumn{1}{r}{$\phantom{0}40${\tiny $\,\%$}} \\
 & \nopagebreak 4.4.1  & \multicolumn{1}{r}{$\phantom{0}14${\tiny $\,\%$}} & \multicolumn{1}{r}{$\phantom{0}86${\tiny $\,\%$}} & \multicolumn{1}{r}{$\phantom{0}43${\tiny $\,\%$}} & \multicolumn{1}{r}{$\phantom{0}69${\tiny $\,\%$}} & \multicolumn{1}{r}{$\phantom{0}48${\tiny $\,\%$}} & \multicolumn{1}{r}{$\phantom{0}12${\tiny $\,\%$}} & \multicolumn{1}{r}{$\phantom{0}90${\tiny $\,\%$}} & \multicolumn{1}{r}{$\phantom{0}12${\tiny $\,\%$}} & \multicolumn{1}{r}{$\phantom{0}14${\tiny $\,\%$}} & \multicolumn{1}{r}{$\phantom{0}40${\tiny $\,\%$}} \\
 & \nopagebreak 4.5.0  & \multicolumn{1}{r}{$\phantom{0}14${\tiny $\,\%$}} & \multicolumn{1}{r}{$\phantom{0}86${\tiny $\,\%$}} & \multicolumn{1}{r}{$\phantom{0}43${\tiny $\,\%$}} & \multicolumn{1}{r}{$\phantom{0}69${\tiny $\,\%$}} & \multicolumn{1}{r}{$\phantom{0}48${\tiny $\,\%$}} & \multicolumn{1}{r}{$\phantom{0}12${\tiny $\,\%$}} & \multicolumn{1}{r}{$\phantom{0}90${\tiny $\,\%$}} & \multicolumn{1}{r}{$\phantom{0}12${\tiny $\,\%$}} & \multicolumn{1}{r}{$\phantom{0}14${\tiny $\,\%$}} & \multicolumn{1}{r}{$\phantom{0}40${\tiny $\,\%$}} \\
 & \nopagebreak \textsc{cvc4}  & \multicolumn{1}{r}{--} & \multicolumn{1}{r}{--} & \multicolumn{1}{r}{--} & \multicolumn{1}{r}{--} & \multicolumn{1}{r}{--} & \multicolumn{1}{r}{--} & \multicolumn{1}{r}{--} & \multicolumn{1}{r}{--} & \multicolumn{1}{r}{--} & \multicolumn{1}{r}{--} \\
 & \nopagebreak \textsc{why3}  & \multicolumn{1}{r}{--} & \multicolumn{1}{r}{--} & \multicolumn{1}{r}{--} & \multicolumn{1}{r}{--} & \multicolumn{1}{r}{--} & \multicolumn{1}{r}{--} & \multicolumn{1}{r}{--} & \multicolumn{1}{r}{--} & \multicolumn{1}{r}{--} & \multicolumn{1}{r}{--} \\
\rule{0pt}{0.9\normalbaselineskip}D & \nopagebreak 4.1.1  & \multicolumn{1}{r}{--} & \multicolumn{1}{r}{--} & \multicolumn{1}{r}{--} & \multicolumn{1}{r}{--} & \multicolumn{1}{r}{--} & \multicolumn{1}{r}{--} & \multicolumn{1}{r}{--} & \multicolumn{1}{r}{--} & \multicolumn{1}{r}{--} & \multicolumn{1}{r}{--} \\
 & \nopagebreak 4.3.2  & \multicolumn{1}{r}{$\phantom{00}0${\tiny $\,\%$}} & \multicolumn{1}{r}{$\phantom{00}0${\tiny $\,\%$}} & \multicolumn{1}{r}{$\phantom{0}56${\tiny $\,\%$}} & \multicolumn{1}{r}{$\phantom{0}63${\tiny $\,\%$}} & \multicolumn{1}{r}{$\phantom{0}80${\tiny $\,\%$}} & \multicolumn{1}{r}{$\phantom{00}0${\tiny $\,\%$}} & \multicolumn{1}{r}{$\phantom{0}70${\tiny $\,\%$}} & \multicolumn{1}{r}{$\phantom{00}0${\tiny $\,\%$}} & \multicolumn{1}{r}{$\phantom{0}68${\tiny $\,\%$}} & \multicolumn{1}{r}{$\phantom{0}41${\tiny $\,\%$}} \\
 & \nopagebreak 4.4.1  & \multicolumn{1}{r}{$\phantom{00}0${\tiny $\,\%$}} & \multicolumn{1}{r}{$\phantom{00}0${\tiny $\,\%$}} & \multicolumn{1}{r}{$\phantom{0}56${\tiny $\,\%$}} & \multicolumn{1}{r}{$\phantom{0}62${\tiny $\,\%$}} & \multicolumn{1}{r}{$\phantom{0}80${\tiny $\,\%$}} & \multicolumn{1}{r}{$\phantom{00}0${\tiny $\,\%$}} & \multicolumn{1}{r}{$\phantom{0}70${\tiny $\,\%$}} & \multicolumn{1}{r}{$\phantom{00}0${\tiny $\,\%$}} & \multicolumn{1}{r}{$\phantom{0}69${\tiny $\,\%$}} & \multicolumn{1}{r}{$\phantom{0}41${\tiny $\,\%$}} \\
 & \nopagebreak 4.5.0  & \multicolumn{1}{r}{$\phantom{00}0${\tiny $\,\%$}} & \multicolumn{1}{r}{$\phantom{00}0${\tiny $\,\%$}} & \multicolumn{1}{r}{$\phantom{0}60${\tiny $\,\%$}} & \multicolumn{1}{r}{$\phantom{0}62${\tiny $\,\%$}} & \multicolumn{1}{r}{$\phantom{0}77${\tiny $\,\%$}} & \multicolumn{1}{r}{$\phantom{00}0${\tiny $\,\%$}} & \multicolumn{1}{r}{$\phantom{0}64${\tiny $\,\%$}} & \multicolumn{1}{r}{$\phantom{00}0${\tiny $\,\%$}} & \multicolumn{1}{r}{$\phantom{0}61${\tiny $\,\%$}} & \multicolumn{1}{r}{$\phantom{0}49${\tiny $\,\%$}} \\
 & \nopagebreak \textsc{cvc4}  & \multicolumn{1}{r}{--} & \multicolumn{1}{r}{--} & \multicolumn{1}{r}{--} & \multicolumn{1}{r}{--} & \multicolumn{1}{r}{--} & \multicolumn{1}{r}{--} & \multicolumn{1}{r}{--} & \multicolumn{1}{r}{--} & \multicolumn{1}{r}{--} & \multicolumn{1}{r}{--} \\
 & \nopagebreak \textsc{why3}  & \multicolumn{1}{r}{--} & \multicolumn{1}{r}{--} & \multicolumn{1}{r}{--} & \multicolumn{1}{r}{--} & \multicolumn{1}{r}{--} & \multicolumn{1}{r}{--} & \multicolumn{1}{r}{--} & \multicolumn{1}{r}{--} & \multicolumn{1}{r}{--} & \multicolumn{1}{r}{--} \\
\rule{0pt}{0.9\normalbaselineskip}E & \nopagebreak 4.1.1  & \multicolumn{1}{r}{--} & \multicolumn{1}{r}{--} & \multicolumn{1}{r}{--} & \multicolumn{1}{r}{--} & \multicolumn{1}{r}{--} & \multicolumn{1}{r}{--} & \multicolumn{1}{r}{--} & \multicolumn{1}{r}{--} & \multicolumn{1}{r}{--} & \multicolumn{1}{r}{--} \\
 & \nopagebreak 4.3.2  & \multicolumn{1}{r}{--} & \multicolumn{1}{r}{--} & \multicolumn{1}{r}{--} & \multicolumn{1}{r}{--} & \multicolumn{1}{r}{--} & \multicolumn{1}{r}{--} & \multicolumn{1}{r}{--} & \multicolumn{1}{r}{--} & \multicolumn{1}{r}{--} & \multicolumn{1}{r}{--} \\
 & \nopagebreak 4.4.1  & \multicolumn{1}{r}{--} & \multicolumn{1}{r}{--} & \multicolumn{1}{r}{--} & \multicolumn{1}{r}{--} & \multicolumn{1}{r}{--} & \multicolumn{1}{r}{--} & \multicolumn{1}{r}{--} & \multicolumn{1}{r}{--} & \multicolumn{1}{r}{--} & \multicolumn{1}{r}{--} \\
 & \nopagebreak 4.5.0  & \multicolumn{1}{r}{--} & \multicolumn{1}{r}{--} & \multicolumn{1}{r}{--} & \multicolumn{1}{r}{--} & \multicolumn{1}{r}{--} & \multicolumn{1}{r}{--} & \multicolumn{1}{r}{--} & \multicolumn{1}{r}{--} & \multicolumn{1}{r}{--} & \multicolumn{1}{r}{--} \\
 & \nopagebreak \textsc{cvc4}  & \multicolumn{1}{r}{--} & \multicolumn{1}{r}{--} & \multicolumn{1}{r}{--} & \multicolumn{1}{r}{--} & \multicolumn{1}{r}{--} & \multicolumn{1}{r}{--} & \multicolumn{1}{r}{--} & \multicolumn{1}{r}{--} & \multicolumn{1}{r}{--} & \multicolumn{1}{r}{--} \\
 & \nopagebreak \textsc{why3}  & \multicolumn{1}{r}{--} & \multicolumn{1}{r}{--} & \multicolumn{1}{r}{--} & \multicolumn{1}{r}{--} & \multicolumn{1}{r}{--} & \multicolumn{1}{r}{--} & \multicolumn{1}{r}{--} & \multicolumn{1}{r}{--} & \multicolumn{1}{r}{--} & \multicolumn{1}{r}{--} \\
\rule{0pt}{0.9\normalbaselineskip}P & \nopagebreak 4.1.1  & \multicolumn{1}{r}{--} & \multicolumn{1}{r}{--} & \multicolumn{1}{r}{--} & \multicolumn{1}{r}{--} & \multicolumn{1}{r}{--} & \multicolumn{1}{r}{--} & \multicolumn{1}{r}{--} & \multicolumn{1}{r}{--} & \multicolumn{1}{r}{--} & \multicolumn{1}{r}{--} \\
 & \nopagebreak 4.3.2  & \multicolumn{1}{r}{$\phantom{00}0${\tiny $\,\%$}} & \multicolumn{1}{r}{$\phantom{00}0${\tiny $\,\%$}} & \multicolumn{1}{r}{$\phantom{0}38${\tiny $\,\%$}} & \multicolumn{1}{r}{$\phantom{0}18${\tiny $\,\%$}} & \multicolumn{1}{r}{$\phantom{0}64${\tiny $\,\%$}} & \multicolumn{1}{r}{$\phantom{00}0${\tiny $\,\%$}} & \multicolumn{1}{r}{$\phantom{0}26${\tiny $\,\%$}} & \multicolumn{1}{r}{$\phantom{00}0${\tiny $\,\%$}} & \multicolumn{1}{r}{$\phantom{0}88${\tiny $\,\%$}} & \multicolumn{1}{r}{$\phantom{0}49${\tiny $\,\%$}} \\
 & \nopagebreak 4.4.1  & \multicolumn{1}{r}{$\phantom{00}0${\tiny $\,\%$}} & \multicolumn{1}{r}{$\phantom{00}0${\tiny $\,\%$}} & \multicolumn{1}{r}{$\phantom{0}30${\tiny $\,\%$}} & \multicolumn{1}{r}{$\phantom{0}17${\tiny $\,\%$}} & \multicolumn{1}{r}{$\phantom{0}57${\tiny $\,\%$}} & \multicolumn{1}{r}{$\phantom{00}0${\tiny $\,\%$}} & \multicolumn{1}{r}{$\phantom{0}27${\tiny $\,\%$}} & \multicolumn{1}{r}{$\phantom{00}0${\tiny $\,\%$}} & \multicolumn{1}{r}{$100${\tiny $\,\%$}} & \multicolumn{1}{r}{$\phantom{0}43${\tiny $\,\%$}} \\
 & \nopagebreak 4.5.0  & \multicolumn{1}{r}{$\phantom{00}0${\tiny $\,\%$}} & \multicolumn{1}{r}{$\phantom{00}0${\tiny $\,\%$}} & \multicolumn{1}{r}{$\phantom{0}30${\tiny $\,\%$}} & \multicolumn{1}{r}{$\phantom{0}17${\tiny $\,\%$}} & \multicolumn{1}{r}{$\phantom{0}57${\tiny $\,\%$}} & \multicolumn{1}{r}{$\phantom{00}0${\tiny $\,\%$}} & \multicolumn{1}{r}{$\phantom{0}27${\tiny $\,\%$}} & \multicolumn{1}{r}{$\phantom{00}0${\tiny $\,\%$}} & \multicolumn{1}{r}{$100${\tiny $\,\%$}} & \multicolumn{1}{r}{$\phantom{0}43${\tiny $\,\%$}} \\
 & \nopagebreak \textsc{cvc4}  & \multicolumn{1}{r}{--} & \multicolumn{1}{r}{--} & \multicolumn{1}{r}{--} & \multicolumn{1}{r}{--} & \multicolumn{1}{r}{--} & \multicolumn{1}{r}{--} & \multicolumn{1}{r}{--} & \multicolumn{1}{r}{--} & \multicolumn{1}{r}{--} & \multicolumn{1}{r}{--} \\
 & \nopagebreak \textsc{why3}  & \multicolumn{1}{r}{--} & \multicolumn{1}{r}{--} & \multicolumn{1}{r}{--} & \multicolumn{1}{r}{--} & \multicolumn{1}{r}{--} & \multicolumn{1}{r}{--} & \multicolumn{1}{r}{--} & \multicolumn{1}{r}{--} & \multicolumn{1}{r}{--} & \multicolumn{1}{r}{--} \\
\rule{0pt}{0.9\normalbaselineskip}S & \nopagebreak 4.1.1  & \multicolumn{1}{r}{--} & \multicolumn{1}{r}{--} & \multicolumn{1}{r}{--} & \multicolumn{1}{r}{--} & \multicolumn{1}{r}{--} & \multicolumn{1}{r}{--} & \multicolumn{1}{r}{--} & \multicolumn{1}{r}{--} & \multicolumn{1}{r}{--} & \multicolumn{1}{r}{--} \\
 & \nopagebreak 4.3.2  & \multicolumn{1}{r}{--} & \multicolumn{1}{r}{--} & \multicolumn{1}{r}{--} & \multicolumn{1}{r}{--} & \multicolumn{1}{r}{--} & \multicolumn{1}{r}{--} & \multicolumn{1}{r}{--} & \multicolumn{1}{r}{--} & \multicolumn{1}{r}{--} & \multicolumn{1}{r}{--} \\
 & \nopagebreak 4.4.1  & \multicolumn{1}{r}{--} & \multicolumn{1}{r}{--} & \multicolumn{1}{r}{--} & \multicolumn{1}{r}{--} & \multicolumn{1}{r}{--} & \multicolumn{1}{r}{--} & \multicolumn{1}{r}{--} & \multicolumn{1}{r}{--} & \multicolumn{1}{r}{--} & \multicolumn{1}{r}{--} \\
 & \nopagebreak 4.5.0  & \multicolumn{1}{r}{--} & \multicolumn{1}{r}{--} & \multicolumn{1}{r}{--} & \multicolumn{1}{r}{--} & \multicolumn{1}{r}{--} & \multicolumn{1}{r}{--} & \multicolumn{1}{r}{--} & \multicolumn{1}{r}{--} & \multicolumn{1}{r}{--} & \multicolumn{1}{r}{--} \\
 & \nopagebreak \textsc{cvc4}  & \multicolumn{1}{r}{--} & \multicolumn{1}{r}{--} & \multicolumn{1}{r}{--} & \multicolumn{1}{r}{--} & \multicolumn{1}{r}{--} & \multicolumn{1}{r}{--} & \multicolumn{1}{r}{--} & \multicolumn{1}{r}{--} & \multicolumn{1}{r}{--} & \multicolumn{1}{r}{--} \\
 & \nopagebreak \textsc{why3}  & \multicolumn{1}{r}{$\phantom{00}5${\tiny $\,\%$}} & \multicolumn{1}{r}{$\phantom{0}32${\tiny $\,\%$}} & \multicolumn{1}{r}{$\phantom{0}44${\tiny $\,\%$}} & \multicolumn{1}{r}{$\phantom{0}52${\tiny $\,\%$}} & \multicolumn{1}{r}{$\phantom{00}0${\tiny $\,\%$}} & \multicolumn{1}{r}{$100${\tiny $\,\%$}} & \multicolumn{1}{r}{$\phantom{0}59${\tiny $\,\%$}} & \multicolumn{1}{r}{$\phantom{0}28${\tiny $\,\%$}} & \multicolumn{1}{r}{$\phantom{0}29${\tiny $\,\%$}} & \multicolumn{1}{r}{$\phantom{0}33${\tiny $\,\%$}} \\
\rule{0pt}{0.9\normalbaselineskip}T & \nopagebreak 4.1.1  & \multicolumn{1}{r}{$\phantom{0}10${\tiny $\,\%$}} & \multicolumn{1}{r}{$\phantom{0}26${\tiny $\,\%$}} & \multicolumn{1}{r}{$\phantom{0}18${\tiny $\,\%$}} & \multicolumn{1}{r}{$\phantom{0}92${\tiny $\,\%$}} & \multicolumn{1}{r}{$\phantom{0}28${\tiny $\,\%$}} & \multicolumn{1}{r}{$\phantom{00}0${\tiny $\,\%$}} & \multicolumn{1}{r}{$\phantom{0}44${\tiny $\,\%$}} & \multicolumn{1}{r}{$\phantom{0}26${\tiny $\,\%$}} & \multicolumn{1}{r}{$\phantom{0}10${\tiny $\,\%$}} & \multicolumn{1}{r}{$\phantom{0}51${\tiny $\,\%$}} \\
 & \nopagebreak 4.3.2  & \multicolumn{1}{r}{$\phantom{0}10${\tiny $\,\%$}} & \multicolumn{1}{r}{$\phantom{0}25${\tiny $\,\%$}} & \multicolumn{1}{r}{$\phantom{0}21${\tiny $\,\%$}} & \multicolumn{1}{r}{$\phantom{0}90${\tiny $\,\%$}} & \multicolumn{1}{r}{$\phantom{0}27${\tiny $\,\%$}} & \multicolumn{1}{r}{$\phantom{00}0${\tiny $\,\%$}} & \multicolumn{1}{r}{$\phantom{0}41${\tiny $\,\%$}} & \multicolumn{1}{r}{$\phantom{0}30${\tiny $\,\%$}} & \multicolumn{1}{r}{$\phantom{0}17${\tiny $\,\%$}} & \multicolumn{1}{r}{$\phantom{0}60${\tiny $\,\%$}} \\
 & \nopagebreak 4.4.1  & \multicolumn{1}{r}{$\phantom{00}8${\tiny $\,\%$}} & \multicolumn{1}{r}{$\phantom{0}26${\tiny $\,\%$}} & \multicolumn{1}{r}{$\phantom{0}21${\tiny $\,\%$}} & \multicolumn{1}{r}{$\phantom{0}90${\tiny $\,\%$}} & \multicolumn{1}{r}{$\phantom{0}28${\tiny $\,\%$}} & \multicolumn{1}{r}{$\phantom{00}0${\tiny $\,\%$}} & \multicolumn{1}{r}{$\phantom{0}41${\tiny $\,\%$}} & \multicolumn{1}{r}{$\phantom{0}31${\tiny $\,\%$}} & \multicolumn{1}{r}{$\phantom{0}18${\tiny $\,\%$}} & \multicolumn{1}{r}{$\phantom{0}61${\tiny $\,\%$}} \\
 & \nopagebreak 4.5.0  & \multicolumn{1}{r}{$\phantom{0}11${\tiny $\,\%$}} & \multicolumn{1}{r}{$\phantom{0}27${\tiny $\,\%$}} & \multicolumn{1}{r}{$\phantom{0}25${\tiny $\,\%$}} & \multicolumn{1}{r}{$\phantom{0}74${\tiny $\,\%$}} & \multicolumn{1}{r}{$\phantom{0}20${\tiny $\,\%$}} & \multicolumn{1}{r}{$\phantom{00}0${\tiny $\,\%$}} & \multicolumn{1}{r}{$\phantom{0}47${\tiny $\,\%$}} & \multicolumn{1}{r}{$\phantom{0}30${\tiny $\,\%$}} & \multicolumn{1}{r}{$\phantom{0}28${\tiny $\,\%$}} & \multicolumn{1}{r}{$\phantom{0}59${\tiny $\,\%$}} \\
 & \nopagebreak \textsc{cvc4}  & \multicolumn{1}{r}{--} & \multicolumn{1}{r}{--} & \multicolumn{1}{r}{--} & \multicolumn{1}{r}{--} & \multicolumn{1}{r}{--} & \multicolumn{1}{r}{--} & \multicolumn{1}{r}{--} & \multicolumn{1}{r}{--} & \multicolumn{1}{r}{--} & \multicolumn{1}{r}{--} \\
 & \nopagebreak \textsc{why3}  & \multicolumn{1}{r}{--} & \multicolumn{1}{r}{--} & \multicolumn{1}{r}{--} & \multicolumn{1}{r}{--} & \multicolumn{1}{r}{--} & \multicolumn{1}{r}{--} & \multicolumn{1}{r}{--} & \multicolumn{1}{r}{--} & \multicolumn{1}{r}{--} & \multicolumn{1}{r}{--} \\
\cmidrule(){2-12}
\rule{0pt}{0.9\normalbaselineskip}\textsl{all} & \nopagebreak 4.1.1  & \multicolumn{1}{r}{$\phantom{00}8${\tiny $\,\%$}} & \multicolumn{1}{r}{$\phantom{0}37${\tiny $\,\%$}} & \multicolumn{1}{r}{$\phantom{0}30${\tiny $\,\%$}} & \multicolumn{1}{r}{$\phantom{0}69${\tiny $\,\%$}} & \multicolumn{1}{r}{$\phantom{0}32${\tiny $\,\%$}} & \multicolumn{1}{r}{$\phantom{00}4${\tiny $\,\%$}} & \multicolumn{1}{r}{$\phantom{0}75${\tiny $\,\%$}} & \multicolumn{1}{r}{$\phantom{0}31${\tiny $\,\%$}} & \multicolumn{1}{r}{$\phantom{0}15${\tiny $\,\%$}} & \multicolumn{1}{r}{$\phantom{0}42${\tiny $\,\%$}} \\
 & \nopagebreak 4.3.2  & \multicolumn{1}{r}{$\phantom{00}2${\tiny $\,\%$}} & \multicolumn{1}{r}{$\phantom{00}7${\tiny $\,\%$}} & \multicolumn{1}{r}{$\phantom{0}45${\tiny $\,\%$}} & \multicolumn{1}{r}{$\phantom{0}47${\tiny $\,\%$}} & \multicolumn{1}{r}{$\phantom{0}68${\tiny $\,\%$}} & \multicolumn{1}{r}{$\phantom{00}1${\tiny $\,\%$}} & \multicolumn{1}{r}{$\phantom{0}52${\tiny $\,\%$}} & \multicolumn{1}{r}{$\phantom{00}3${\tiny $\,\%$}} & \multicolumn{1}{r}{$\phantom{0}69${\tiny $\,\%$}} & \multicolumn{1}{r}{$\phantom{0}45${\tiny $\,\%$}} \\
 & \nopagebreak 4.4.1  & \multicolumn{1}{r}{$\phantom{00}2${\tiny $\,\%$}} & \multicolumn{1}{r}{$\phantom{00}8${\tiny $\,\%$}} & \multicolumn{1}{r}{$\phantom{0}43${\tiny $\,\%$}} & \multicolumn{1}{r}{$\phantom{0}49${\tiny $\,\%$}} & \multicolumn{1}{r}{$\phantom{0}66${\tiny $\,\%$}} & \multicolumn{1}{r}{$\phantom{00}1${\tiny $\,\%$}} & \multicolumn{1}{r}{$\phantom{0}54${\tiny $\,\%$}} & \multicolumn{1}{r}{$\phantom{00}3${\tiny $\,\%$}} & \multicolumn{1}{r}{$\phantom{0}72${\tiny $\,\%$}} & \multicolumn{1}{r}{$\phantom{0}43${\tiny $\,\%$}} \\
 & \nopagebreak 4.5.0  & \multicolumn{1}{r}{$\phantom{00}2${\tiny $\,\%$}} & \multicolumn{1}{r}{$\phantom{00}9${\tiny $\,\%$}} & \multicolumn{1}{r}{$\phantom{0}44${\tiny $\,\%$}} & \multicolumn{1}{r}{$\phantom{0}46${\tiny $\,\%$}} & \multicolumn{1}{r}{$\phantom{0}62${\tiny $\,\%$}} & \multicolumn{1}{r}{$\phantom{00}1${\tiny $\,\%$}} & \multicolumn{1}{r}{$\phantom{0}50${\tiny $\,\%$}} & \multicolumn{1}{r}{$\phantom{00}3${\tiny $\,\%$}} & \multicolumn{1}{r}{$\phantom{0}70${\tiny $\,\%$}} & \multicolumn{1}{r}{$\phantom{0}47${\tiny $\,\%$}} \\
 & \nopagebreak \textsc{cvc4}  & \multicolumn{1}{r}{--} & \multicolumn{1}{r}{--} & \multicolumn{1}{r}{--} & \multicolumn{1}{r}{--} & \multicolumn{1}{r}{--} & \multicolumn{1}{r}{--} & \multicolumn{1}{r}{--} & \multicolumn{1}{r}{--} & \multicolumn{1}{r}{--} & \multicolumn{1}{r}{--} \\
 & \nopagebreak \textsc{why3}  & \multicolumn{1}{r}{$\phantom{00}5${\tiny $\,\%$}} & \multicolumn{1}{r}{$\phantom{0}32${\tiny $\,\%$}} & \multicolumn{1}{r}{$\phantom{0}44${\tiny $\,\%$}} & \multicolumn{1}{r}{$\phantom{0}52${\tiny $\,\%$}} & \multicolumn{1}{r}{$\phantom{00}0${\tiny $\,\%$}} & \multicolumn{1}{r}{$100${\tiny $\,\%$}} & \multicolumn{1}{r}{$\phantom{0}59${\tiny $\,\%$}} & \multicolumn{1}{r}{$\phantom{0}28${\tiny $\,\%$}} & \multicolumn{1}{r}{$\phantom{0}29${\tiny $\,\%$}} & \multicolumn{1}{r}{$\phantom{0}33${\tiny $\,\%$}} \\
\bottomrule 
\end{tabular}
  
\caption{
\scriptsize
{Analysis of mutation operators in batch $M_*$.}
{
For each \textsc{group} of seeds, for each \textsc{tool},
for each mutation operator OP, 
column OP reports the percentage of \emph{failing mutants of passing seeds} that were generated
by applying one or more times operator OP (possibly in combination with other mutation operators).
}
}
\label{tab:muops}
\end{table}
\fi

\pparagraph{Failures.}
Overall, 13 brittle seeds are revealed by 350 failing mutants in $M_*$ with Boogie~4.5.0.
Failures are of three kinds:
\begin{enumerate*}[label=\emph{\alph*)}]
\item timeouts (6 seeds, 252 mutants);
\item type errors (5 seeds, 10 mutants);
\item explicit verification failures (2 seeds, 88 mutants).
\end{enumerate*}
\iflong
\begin{description}
\item[\emph{Timeouts}]
\else \emph{Timeouts} \fi
mainly occur in group D (5 seeds), where size and complexity of the code are such that any mutation that slows down verification may hit the timeout limit; verification of some mutants seems to be non-terminating, whereas others are just slowed down by some tens of seconds.
\iflong
One exception is \B{BQueue} in group T, whose implementation of a queue in the style of dynamic frames is not particularly large (322 lines) but includes many assertions that take time to verify.
Some mutants verify if given longer time; in fact, group T's programs are otherwise very robust, probably because they are part of Boogie's test suite, and thus any change in Boogie is checked against the same examples to ensure they still verify.
\fi
\iflong
\item[\emph{Type errors}]
\else \emph{Type errors} \fi all occur in group P and only when mutation \ref{m:S6} splits the seed in a way that procedure \B{update_heap} (part of AutoProof's heap axiomatization) ends up being declared after its first usage; in this case, Boogie cannot correctly instantiate the procedure's generic type, which triggers a type error even before Z3 is involved.
\iflong
Even though AutoProof's heap encoding is based on Dafny's and hence somewhat similar to it,
Dafny is immune to such faulty behavior.
\fi
\iflong
\item[\emph{Verification failures}]
\else \emph{Verification failures} \fi occur in seeds of group A and D.
In particular, 
a binary search tree implementation in group A fails verification when the relative order of two postconditions is swapped by \ref{m:L6};
while Why3 cannot prove the whole example, it can prove the brittle procedure alone regardless of the postcondition order.
\iflong
Group D's solution to problem~3 in the VerifyThis 2015 competition~\cite{VT15} fails verification when two preconditions are merged into a conjunctive one by \ref{m:L4}.
\fi
\iflong
\end{description}
\fi
In all, it is clear that Boogie's encoding is quite sensitive to the order of declarations and assertions even when it should not matter.

\pparagraph{Triggers.}
Remember that mutation operator \ref{m:G2} is the only one that modifies triggers, and was only applied in isolation in a separate set of experiments.
As we expected from previous work~\cite{triggerSelection}, altering triggers is likely to make verification fail (30 seeds and 276 mutants overall; 20 seeds are only brittle if triggers are modified);
most of these failures (26 seeds and 250 mutants) are timeouts, since removing triggers is likely to at least slow down verification---if not make it diverge.
Operator \ref{m:G2} is very effective at exposing brittleness mainly with the complex examples in groups D and P, which include numerous axioms and extensive quantification patterns.
Group E's programs are a bit special because they are brittle---they are designed to be so---but are only affected by mutation operators that remove the trigger annotations on which they strongly depend; in contrast, they are robust against all other mutation operators.

\iflong
\pparagraph{Failing seeds.}
\autoref{tab:pass} is the counterpart of \autoref{tab:fail}, showing if random mutations may change a seed that fails verification into one that passes it.
Unsurprisingly, this does not happen very often:
there are only 2 seeds that go from failing to passing with random mutations.
One in group E is similar to the example of \autoref{sec:example} but where the seed's order of declarations fails verification, and swapping two of them restores verifiability;
one in group P is \B{sum_and_max.bpl}, which robustly verifies with Boogie~4.3.2 but times out with more recent Boogie versions.
AutoProof's encoding was fine-tuned based on Boogie 4.3.2, which explains why it may be sensitive to using newer Boogie versions.

\begin{table}[!tbh]
\centering
\scriptsize
\begin{tabular}{llccccc}
\toprule
& &
\multicolumn{5}{c}{\textsc{batch} $M_*$}
\\
\cmidrule(lr){3-7}
\textsc{group} & \textsc{tool} & \textsc{\# fail} & \textsc{\# $\exists$pass} & \textsc{\% $\exists$pass} & $\overline{\textsc{\% pass}}$ & \multicolumn{1}{c}{$\overline{\textsc{\% }\exists\textsc{pass}}$} \\ 
\midrule
\nopagebreak A & \nopagebreak 4.1.1  & \multicolumn{1}{r}{$\phantom{0}0$} & \multicolumn{1}{r}{$0$} & \multicolumn{1}{r}{--} & \multicolumn{1}{r}{--} & \multicolumn{1}{r}{--} \\
 & \nopagebreak 4.3.2  & \multicolumn{1}{r}{$\phantom{0}0$} & \multicolumn{1}{r}{$0$} & \multicolumn{1}{r}{--} & \multicolumn{1}{r}{--} & \multicolumn{1}{r}{--} \\
 & \nopagebreak 4.4.1  & \multicolumn{1}{r}{$\phantom{0}0$} & \multicolumn{1}{r}{$0$} & \multicolumn{1}{r}{--} & \multicolumn{1}{r}{--} & \multicolumn{1}{r}{--} \\
 & \nopagebreak 4.5.0  & \multicolumn{1}{r}{$\phantom{0}0$} & \multicolumn{1}{r}{$0$} & \multicolumn{1}{r}{--} & \multicolumn{1}{r}{--} & \multicolumn{1}{r}{--} \\
 & \nopagebreak \textsc{cvc4}  & \multicolumn{1}{r}{$\phantom{0}4$} & \multicolumn{1}{r}{$0$} & \multicolumn{1}{r}{$\phantom{00}0${\tiny $\,\%$}} & \multicolumn{1}{r}{$\phantom{00}0${\tiny $\,\%$}} & \multicolumn{1}{r}{--} \\
 & \nopagebreak \textsc{why3}  & \multicolumn{1}{r}{$\phantom{0}3$} & \multicolumn{1}{r}{$0$} & \multicolumn{1}{r}{$\phantom{00}0${\tiny $\,\%$}} & \multicolumn{1}{r}{$\phantom{00}0${\tiny $\,\%$}} & \multicolumn{1}{r}{--} \\
\rule{0pt}{0.9\normalbaselineskip}D & \nopagebreak 4.1.1  & \multicolumn{1}{r}{$26$} & \multicolumn{1}{r}{$0$} & \multicolumn{1}{r}{$\phantom{00}0${\tiny $\,\%$}} & \multicolumn{1}{r}{$\phantom{00}0${\tiny $\,\%$}} & \multicolumn{1}{r}{--} \\
 & \nopagebreak 4.3.2  & \multicolumn{1}{r}{$\phantom{0}2$} & \multicolumn{1}{r}{$0$} & \multicolumn{1}{r}{$\phantom{00}0${\tiny $\,\%$}} & \multicolumn{1}{r}{$\phantom{00}0${\tiny $\,\%$}} & \multicolumn{1}{r}{--} \\
 & \nopagebreak 4.4.1  & \multicolumn{1}{r}{$\phantom{0}2$} & \multicolumn{1}{r}{$0$} & \multicolumn{1}{r}{$\phantom{00}0${\tiny $\,\%$}} & \multicolumn{1}{r}{$\phantom{00}0${\tiny $\,\%$}} & \multicolumn{1}{r}{--} \\
 & \nopagebreak 4.5.0  & \multicolumn{1}{r}{$\phantom{0}2$} & \multicolumn{1}{r}{$0$} & \multicolumn{1}{r}{$\phantom{00}0${\tiny $\,\%$}} & \multicolumn{1}{r}{$\phantom{00}0${\tiny $\,\%$}} & \multicolumn{1}{r}{--} \\
 & \nopagebreak \textsc{cvc4}  & \multicolumn{1}{r}{$26$} & \multicolumn{1}{r}{$0$} & \multicolumn{1}{r}{$\phantom{00}0${\tiny $\,\%$}} & \multicolumn{1}{r}{$\phantom{00}0${\tiny $\,\%$}} & \multicolumn{1}{r}{--} \\
 & \nopagebreak \textsc{why3}  & \multicolumn{1}{r}{$26$} & \multicolumn{1}{r}{$0$} & \multicolumn{1}{r}{$\phantom{00}0${\tiny $\,\%$}} & \multicolumn{1}{r}{$\phantom{00}0${\tiny $\,\%$}} & \multicolumn{1}{r}{--} \\
\rule{0pt}{0.9\normalbaselineskip}E & \nopagebreak 4.1.1  & \multicolumn{1}{r}{$\phantom{0}5$} & \multicolumn{1}{r}{$0$} & \multicolumn{1}{r}{$\phantom{00}0${\tiny $\,\%$}} & \multicolumn{1}{r}{$\phantom{00}0${\tiny $\,\%$}} & \multicolumn{1}{r}{--} \\
 & \nopagebreak 4.3.2  & \multicolumn{1}{r}{$\phantom{0}3$} & \multicolumn{1}{r}{$1$} & \multicolumn{1}{r}{$\phantom{0}33${\tiny $\,\%$}} & \multicolumn{1}{r}{$\phantom{00}6${\tiny $\,\%$}} & \multicolumn{1}{r}{$\phantom{0}17${\tiny $\,\%$}} \\
 & \nopagebreak 4.4.1  & \multicolumn{1}{r}{$\phantom{0}3$} & \multicolumn{1}{r}{$1$} & \multicolumn{1}{r}{$\phantom{0}33${\tiny $\,\%$}} & \multicolumn{1}{r}{$\phantom{00}6${\tiny $\,\%$}} & \multicolumn{1}{r}{$\phantom{0}17${\tiny $\,\%$}} \\
 & \nopagebreak 4.5.0  & \multicolumn{1}{r}{$\phantom{0}3$} & \multicolumn{1}{r}{$1$} & \multicolumn{1}{r}{$\phantom{0}33${\tiny $\,\%$}} & \multicolumn{1}{r}{$\phantom{00}6${\tiny $\,\%$}} & \multicolumn{1}{r}{$\phantom{0}17${\tiny $\,\%$}} \\
 & \nopagebreak \textsc{cvc4}  & \multicolumn{1}{r}{$\phantom{0}6$} & \multicolumn{1}{r}{$0$} & \multicolumn{1}{r}{$\phantom{00}0${\tiny $\,\%$}} & \multicolumn{1}{r}{$\phantom{00}0${\tiny $\,\%$}} & \multicolumn{1}{r}{--} \\
 & \nopagebreak \textsc{why3}  & \multicolumn{1}{r}{$\phantom{0}3$} & \multicolumn{1}{r}{$0$} & \multicolumn{1}{r}{$\phantom{00}0${\tiny $\,\%$}} & \multicolumn{1}{r}{$\phantom{00}0${\tiny $\,\%$}} & \multicolumn{1}{r}{--} \\
\rule{0pt}{0.9\normalbaselineskip}P & \nopagebreak 4.1.1  & \multicolumn{1}{r}{$30$} & \multicolumn{1}{r}{$0$} & \multicolumn{1}{r}{$\phantom{00}0${\tiny $\,\%$}} & \multicolumn{1}{r}{$\phantom{00}0${\tiny $\,\%$}} & \multicolumn{1}{r}{--} \\
 & \nopagebreak 4.3.2  & \multicolumn{1}{r}{$16$} & \multicolumn{1}{r}{$0$} & \multicolumn{1}{r}{$\phantom{00}0${\tiny $\,\%$}} & \multicolumn{1}{r}{$\phantom{00}0${\tiny $\,\%$}} & \multicolumn{1}{r}{--} \\
 & \nopagebreak 4.4.1  & \multicolumn{1}{r}{$17$} & \multicolumn{1}{r}{$1$} & \multicolumn{1}{r}{$\phantom{00}6${\tiny $\,\%$}} & \multicolumn{1}{r}{$\phantom{00}1${\tiny $\,\%$}} & \multicolumn{1}{r}{$\phantom{0}18${\tiny $\,\%$}} \\
 & \nopagebreak 4.5.0  & \multicolumn{1}{r}{$17$} & \multicolumn{1}{r}{$1$} & \multicolumn{1}{r}{$\phantom{00}6${\tiny $\,\%$}} & \multicolumn{1}{r}{$\phantom{00}0${\tiny $\,\%$}} & \multicolumn{1}{r}{$\phantom{00}5${\tiny $\,\%$}} \\
 & \nopagebreak \textsc{cvc4}  & \multicolumn{1}{r}{$30$} & \multicolumn{1}{r}{$0$} & \multicolumn{1}{r}{$\phantom{00}0${\tiny $\,\%$}} & \multicolumn{1}{r}{$\phantom{00}0${\tiny $\,\%$}} & \multicolumn{1}{r}{--} \\
 & \nopagebreak \textsc{why3}  & \multicolumn{1}{r}{$30$} & \multicolumn{1}{r}{$0$} & \multicolumn{1}{r}{$\phantom{00}0${\tiny $\,\%$}} & \multicolumn{1}{r}{$\phantom{00}0${\tiny $\,\%$}} & \multicolumn{1}{r}{--} \\
\rule{0pt}{0.9\normalbaselineskip}S & \nopagebreak 4.1.1  & \multicolumn{1}{r}{$\phantom{0}0$} & \multicolumn{1}{r}{$0$} & \multicolumn{1}{r}{--} & \multicolumn{1}{r}{--} & \multicolumn{1}{r}{--} \\
 & \nopagebreak 4.3.2  & \multicolumn{1}{r}{$\phantom{0}0$} & \multicolumn{1}{r}{$0$} & \multicolumn{1}{r}{--} & \multicolumn{1}{r}{--} & \multicolumn{1}{r}{--} \\
 & \nopagebreak 4.4.1  & \multicolumn{1}{r}{$\phantom{0}0$} & \multicolumn{1}{r}{$0$} & \multicolumn{1}{r}{--} & \multicolumn{1}{r}{--} & \multicolumn{1}{r}{--} \\
 & \nopagebreak 4.5.0  & \multicolumn{1}{r}{$\phantom{0}0$} & \multicolumn{1}{r}{$0$} & \multicolumn{1}{r}{--} & \multicolumn{1}{r}{--} & \multicolumn{1}{r}{--} \\
 & \nopagebreak \textsc{cvc4}  & \multicolumn{1}{r}{$\phantom{0}0$} & \multicolumn{1}{r}{$0$} & \multicolumn{1}{r}{--} & \multicolumn{1}{r}{--} & \multicolumn{1}{r}{--} \\
 & \nopagebreak \textsc{why3}  & \multicolumn{1}{r}{$11$} & \multicolumn{1}{r}{$0$} & \multicolumn{1}{r}{$\phantom{00}0${\tiny $\,\%$}} & \multicolumn{1}{r}{$\phantom{00}0${\tiny $\,\%$}} & \multicolumn{1}{r}{--} \\
\rule{0pt}{0.9\normalbaselineskip}T & \nopagebreak 4.1.1  & \multicolumn{1}{r}{$\phantom{0}0$} & \multicolumn{1}{r}{$0$} & \multicolumn{1}{r}{--} & \multicolumn{1}{r}{--} & \multicolumn{1}{r}{--} \\
 & \nopagebreak 4.3.2  & \multicolumn{1}{r}{$\phantom{0}0$} & \multicolumn{1}{r}{$0$} & \multicolumn{1}{r}{--} & \multicolumn{1}{r}{--} & \multicolumn{1}{r}{--} \\
 & \nopagebreak 4.4.1  & \multicolumn{1}{r}{$\phantom{0}0$} & \multicolumn{1}{r}{$0$} & \multicolumn{1}{r}{--} & \multicolumn{1}{r}{--} & \multicolumn{1}{r}{--} \\
 & \nopagebreak 4.5.0  & \multicolumn{1}{r}{$\phantom{0}0$} & \multicolumn{1}{r}{$0$} & \multicolumn{1}{r}{--} & \multicolumn{1}{r}{--} & \multicolumn{1}{r}{--} \\
 & \nopagebreak \textsc{cvc4}  & \multicolumn{1}{r}{$\phantom{0}4$} & \multicolumn{1}{r}{$0$} & \multicolumn{1}{r}{$\phantom{00}0${\tiny $\,\%$}} & \multicolumn{1}{r}{$\phantom{00}0${\tiny $\,\%$}} & \multicolumn{1}{r}{--} \\
 & \nopagebreak \textsc{why3}  & \multicolumn{1}{r}{$\phantom{0}5$} & \multicolumn{1}{r}{$0$} & \multicolumn{1}{r}{$\phantom{00}0${\tiny $\,\%$}} & \multicolumn{1}{r}{$\phantom{00}0${\tiny $\,\%$}} & \multicolumn{1}{r}{--} \\
\cmidrule(){2-7}
\rule{0pt}{0.9\normalbaselineskip}\textsl{all} & \nopagebreak 4.1.1  & \multicolumn{1}{r}{$61$} & \multicolumn{1}{r}{$0$} & \multicolumn{1}{r}{$\phantom{00}0${\tiny $\,\%$}} & \multicolumn{1}{r}{$\phantom{00}0${\tiny $\,\%$}} & \multicolumn{1}{r}{--} \\
 & \nopagebreak 4.3.2  & \multicolumn{1}{r}{$21$} & \multicolumn{1}{r}{$1$} & \multicolumn{1}{r}{$\phantom{00}5${\tiny $\,\%$}} & \multicolumn{1}{r}{$\phantom{00}1${\tiny $\,\%$}} & \multicolumn{1}{r}{$\phantom{0}17${\tiny $\,\%$}} \\
 & \nopagebreak 4.4.1  & \multicolumn{1}{r}{$22$} & \multicolumn{1}{r}{$2$} & \multicolumn{1}{r}{$\phantom{00}9${\tiny $\,\%$}} & \multicolumn{1}{r}{$\phantom{00}2${\tiny $\,\%$}} & \multicolumn{1}{r}{$\phantom{0}17${\tiny $\,\%$}} \\
 & \nopagebreak 4.5.0  & \multicolumn{1}{r}{$22$} & \multicolumn{1}{r}{$2$} & \multicolumn{1}{r}{$\phantom{00}9${\tiny $\,\%$}} & \multicolumn{1}{r}{$\phantom{00}1${\tiny $\,\%$}} & \multicolumn{1}{r}{$\phantom{0}11${\tiny $\,\%$}} \\
 & \nopagebreak \textsc{cvc4}  & \multicolumn{1}{r}{$70$} & \multicolumn{1}{r}{$0$} & \multicolumn{1}{r}{$\phantom{00}0${\tiny $\,\%$}} & \multicolumn{1}{r}{$\phantom{00}0${\tiny $\,\%$}} & \multicolumn{1}{r}{--} \\
 & \nopagebreak \textsc{why3}  & \multicolumn{1}{r}{$78$} & \multicolumn{1}{r}{$0$} & \multicolumn{1}{r}{$\phantom{00}0${\tiny $\,\%$}} & \multicolumn{1}{r}{$\phantom{00}0${\tiny $\,\%$}} & \multicolumn{1}{r}{--} \\
\bottomrule 
\end{tabular}
  
\caption{
\scriptsize
{
For each \textsc{group} of seeds, for each \textsc{tool}:
number of seeds failing verification (\textsc{\# fail}),
number and percentage of failing seeds for which at least one mutant passes verification (\textsc{\#/\% $\exists$pass}),
 average percentage of mutants per failing seed that pass verification ($\overline{\textsc{\% pass}}$),
average percentage of mutants that pass verification per failing seed with at least one passing mutant ($\overline{\textsc{\% pass}}$).
All data in the table is about experiments with batch $M_*$.
}
}
\label{tab:pass}
\end{table}

\fi

\section{Related Work}
\label{sec:related}

\pparagraph{Robustness.}
This paper's robustness testing aims at detecting
so-called \emph{butterfly effects}~\cite{triggerSelection}---macroscopic 
changes in a verifier's output in response to minor modifications of its input.
Program provers often incur volatile behavior because they use automated theorem provers---such as SMT solvers---which in turn rely on heuristics to handle efficiently, in many practical cases, complex proofs in undecidable logics.
\iflong
Matching triggers---heuristics to guide quantifier instantiation---are 
especially prone to misfire in response to tiny changes in the input, 
as observed in previous work~\cite{triggerSelection,blt} and confirmed by our experiments in \autoref{sec:results}.
\fi

\iflong
The notion of robustness originates from dynamical systems theory~\cite{FMMR-TimeBook-12,Kha95}.
While robustness is well understood for linear systems, \emph{nonlinear} systems may manifest unpredictable loss of robustness that are hard to analyze and prevent.
\fi
\iflong
In this context, real time temporal logics have been proposed as a way of 
formalizing and analyzing behavioral properties that are satisfied robustly~\cite{FainekosP09}.
\fi

\pparagraph{Random testing.}
Our approach uses \emph{testing} to expose brittle behavior of verifiers.
\iflong
While testing can only try out finitely many inputs---and
thus can only prove the presence of errors, as remarked in one of Dijkstra's most memorable quotes\iflong~\cite{dijkstra1970notes}\fi---it is an invaluable
analysis techniques, which requires relatively little effort to be applied.
\fi
By automatically generating test inputs, \emph{random testing} has proved to be extremely effective at detecting subtle errors in programs completely automatically.
Random testing can generate instances of complex data types by recursively building them according to their inductive structure---as it has been done for functional~\iflong\cite{QuickCheck,FetscherCPHF15}\else\cite{QuickCheck}\fi and object-oriented~\iflong\cite{Randoop,AutoTest}\else\cite{Randoop}\fi programming languages.
Random testing has also been successfully applied to security testing---where it is normally called ``fuzzing''~\cite{GodefroidLM12}---as well as to compiler testing~\iflong\cite{csmith,LeAS14}\else\cite{csmith}\fi---where well-formed programs are randomly generated according to the input language's grammar.

\pparagraph{Mutation testing.}
This paper's robustness testing is a form of random testing, in that it applies random mutation operators to transform a program into an equivalent one.
The terminology and the idea of applying mutation operators to transform between variants of a program come from \emph{mutation testing}~\cite{mutationSurvey}.
However, the goals of traditional mutation testing and of this paper's robustness testing are specular.
Mutation testing is normally used to assess the robustness of a test suite---by 
applying error-inducing mutations to correct programs, and ascertaining whether the tests fail on the mutated programs.
In contrast, we use mutation testing to assess the \emph{robustness of a verifier}---by applying semantic-preserving mutations to correct (verified) programs, and ascertaining whether the mutated programs still verify.
Therefore, the mutation operators of standard mutation testing introduce bugs in a way that is representative of common programming mistakes; the mutation operators of robustness testing (see \autoref{tab:transformations}) do not alter correctness but merely represent alternative syntax expressing the same behavior in a way that is representative of different styles of programming.

\pparagraph{Metamorphic testing.}
In testing, generating inputs is only half of the work;
one also has to compare the system's output with the \emph{expected output} to determine whether a test is passing or failing.
The definition of correct expected output is given by an \emph{oracle}~\cite{BarrHMSY15}.
The more complex the properties we are testing for, the more complex the oracle:
a crash oracle (did the program crash?) is sufficient to test for simple errors such as out-of-bound memory access;
finding more complex errors requires some form of \emph{specification}~\cite{HieronsBBCDDGHKKLSVWZ09} of expected behavior\iflong---for example in the form of assertions~\cite{Korat,PFPWM-ICSE13}\fi.

Even when directly building an oracle is as complex as writing a correct program,
there are still indirect ways of extrapolating whether an output is correct.
In \emph{differential testing}~\cite{differentialTesting}, there are variants of the program under test;
under the assumption that not all variants have the same bugs, one can feed the same input to every variant, and stipulate that the output returned by the majority is the expected one---and any outlier is likely buggy.
\iflong
Differential testing has been applied to testing compilers~\cite{csmith}, looking for the compiler that generates the executable that behaves differently from the others on the same input.
\fi
In \emph{metamorphic testing}~\iflong\cite{metamorphic,metamorphicSurvey}\else\cite{metamorphicSurvey}\fi,
an input is transformed into an equivalent one according to \emph{metamorphic relations}\iflong{} (for example, the inputs $x$ and $-x$ should be equivalent inputs to a function computing the absolute value)\fi; equivalent inputs that determine different outputs are indicative of error.
Our robustness testing applies mutation operators that determine identity metamorphic relations between Boogie programs, since they only change syntactic details and not the semantics of programs.

\section{Discussion and Future Work}
\label{sec:discussion}

Our experiments with \mugie confirm the intuition---bred by frequently using it in our work---that Boogie is prone to brittle behaviour.
How can we shield users from this brittle behavior, thus improving the usability of verification technology?

Program verifiers that use Boogie as an intermediate representation achieve this goal to some extent: 
the researchers who built the verifiers have developed an intuitive understanding of Boogie's idiosyncrasies, and have encoded this informal knowledge into their tools.
End users do not have to worry about Boogie's brittleness but can count on the tools to provide an encoding of their input programs that has a good chance of being effective.
In contrast, developers of program verifiers still have to know how to interact with Boogie and be aware of its peculiarities.

Robustness testing may play a role not only in exposing brittle behavior---the focus of this paper---but in precisely tracking down the sources of brittleness, thus helping to debug them.
To this end, we plan to address \emph{minimization} and \emph{equivalency detection} of mutants in future work.
The idea is that the number of failing mutants that we get by running \mugie are not directly effective as debugging aids, because it takes a good deal of manual analysis to pinpoint the precise sources of failure in large programs with several mutations.
Instead, we will apply techniques such as delta debugging~\cite{deltaDebug} to reduce the size of a failing mutant as much as possible while still triggering failing behavior in Boogie.
Failing mutants of minimal size will be easier to inspect by hand, and thus will point to concrete aspects of the Boogie translation that could be made more robust.

To further investigate to what extent it is Z3 that is brittle, and to what extent it is Boogie's encoding of verification condition---an aspect only partially addressed by this paper's experiments---we will apply robustness testing directly to SMT problems, also to understand how Boogie's encoding can be made more robust. 

Robustness testing could become a useful technique for developers working on different layers of verification infrastructure, to help them track down sources of brittleness during development and ultimately making verification technology easier to use and more broadly applicable.

\bibliographystyle{splncs_srt}
\bibliography{robustness}

\end{document}